\pdfoutput=1
\documentclass[11pt,a4paper]{article}
\usepackage{jheppub}

\RequirePackage[utf8]{inputenc}

\usepackage{amsfonts}
\usepackage{amssymb}
\usepackage{comment}
\usepackage{graphicx}
\usepackage{amsmath}
\usepackage{tabu}
\usepackage{dsfont}
\usepackage{float}
\usepackage{amsthm}
\usepackage{slashed}
\usepackage{setspace}
\usepackage[labelformat=simple]{subcaption}

\usepackage{pgfplots}
\usepackage{tikz}
\usepackage{tikz-cd}
\usetikzlibrary{shapes.misc,shadows,fit,decorations.pathmorphing,positioning,trees,decorations.markings,decorations.pathreplacing,calc,shapes,patterns,arrows}
\usepackage{xcolor}
\usepackage[boxsize=0.5em,aligntableaux=center]{ytableau}
\usepackage{makecell}
\usepackage{environ}
\usepackage[utf8]{inputenc}
\usepackage{amsmath, amsthm, amssymb, amsfonts}
\usepackage[font=small, labelfont=bf]{caption}
\usepackage{amsmath, amsthm, amssymb, amsfonts}
\usepackage{multirow}
\usepackage{graphicx}
\usepackage{float}
\usepackage[numbers,sort&compress]{natbib}
\usepackage{jheppub}
\usepackage{enumerate}
\usepackage{hhline}
\usepackage{amsmath}
\usepackage{amsfonts}
\usepackage{amssymb}
\usepackage{graphicx}
\usepackage{cancel}
\usepackage{makecell}
\usepackage{tabularx}
\usepackage{cellspace}
\usepackage{longtable}
\usepackage{mathtools}
\usepackage{layouts}
\usepackage{empheq}
\usepackage{color}
\usepackage{hyperref}
\usepackage[capitalize,sort]{cleveref}

\crefname{figure}{Figure}{Figures}
\crefname{table}{Table}{Tables}

\usepackage{float}
\restylefloat{table}
\restylefloat{figure}

\newcommand{\nc}{\newcommand}
\nc{\beq}{\begin{equation}}
\nc{\eeq}{\end{equation}}
\nc{\bea}{\begin{eqnarray}}
\nc{\eea}{\end{eqnarray}}

\nc{\be}{\begin{equation}}
\nc{\ee}{\end{equation}}

\def\ov{\overline}

\def\vo{\mathcal{V}}

\newcommand{\la}{\langle}
\newcommand{\ra}{\rangle}

\allowdisplaybreaks[1]
\numberwithin{equation}{section}

\newcommand{\CC}{\mathbb{C}}
\newcommand{\PP}{\mathbb{P}}

\newcommand{\coma}{\, , \quad}
\newcommand{\fstop}{\, .}

\renewcommand{\epsilon}{\varepsilon}

\makeatletter
\newsavebox{\measure@tikzpicture}
\NewEnviron{scaletikzpicturetowidth}[1]{%
  \def\tikz@width{#1}%
  \begin{lrbox}{\measure@tikzpicture}%
  \BODY
  \end{lrbox}%
  \pgfmathparse{#1/\wd\measure@tikzpicture}%
  \BODY
}
\makeatother

\catcode`\@=11   
\newdimen\@rotdimen
\newbox\@rotbox  

\def\@vspec#1{\special{ps:#1}}
\def\@rotstart#1{\@vspec{gsave currentpoint currentpoint translate
		#1 neg exch neg exch translate}}
\def\@rotfinish{\@vspec{currentpoint grestore moveto}}
\def\@rotr#1{\@rotdimen=\ht#1\advance\@rotdimen by\dp#1%
	\hbox to\@rotdimen{\hskip\ht#1\vbox to\wd#1{\@rotstart{90 rotate}%
			\box#1\vss}\hss}\@rotfinish}
\def\@rotl#1{\@rotdimen=\ht#1\advance\@rotdimen by\dp#1%
	\hbox to\@rotdimen{\vbox to\wd#1{\vskip\wd#1\@rotstart{270 rotate}%
			\box#1\vss}\hss}\@rotfinish}%
\def\@rotu#1{\@rotdimen=\ht#1\advance\@rotdimen by\dp#1%
	\hbox to\wd#1{\hskip\wd#1\vbox to\@rotdimen{\vskip\@rotdimen
			\@rotstart{-1 dup scale}\box#1\vss}\hss}\@rotfinish}%
\def\@rotf#1{\hbox to\wd#1{\hskip\wd#1\@rotstart{-1 1 scale}%
		\box#1\hss}\@rotfinish}%
\def\rotate{\@ifnextchar[{\@rotate}{\@rotate[l]}}
\def\@rotate[#1]#2{\setbox\@rotbox=\hbox{#2}\@nameuse{@rot#1}\@rotbox}

\catcode`\@=12

\tikzset{
    partial ellipse/.style args={#1:#2:#3}{
        insert path={+ (#1:#3) arc (#1:#2:#3)}
    }
}

\preprint{ZMP-HH/22-2}

\title{Divisor topologies of CICY 3-folds and their applications to phenomenology 
}

\author[a]{Federico Carta,}
\author[b]{Alessandro Mininno,}
\author[c]{Pramod Shukla}

\affiliation[a]{Department of Mathematical Sciences, Durham University, \\ Durham, DH1 3LE, United Kingdom}
\affiliation[b]{II. Institut f\"ur Theoretische Physik, Universit\"at Hamburg,\\
Luruper Chaussee 149, 22607 Hamburg, Germany}
\affiliation[c]{\small ICTP, Strada Costiera 11, Trieste 34151, Italy}

\emailAdd{federico.carta@durham.ac.uk}
\emailAdd{alessandro.mininno@desy.de}
\emailAdd{pramodmaths@gmail.com}

\abstract{In this article, we present a classification for the divisor topologies of the projective complete intersection Calabi-Yau (pCICY) 3-folds realized as hypersurfaces in the product of complex projective spaces. There are 7890 such pCICYs of which 7820 are favorable, and can be subsequently useful for phenomenological purposes. To our surprise we find that the whole pCICY database results in only 11 (so-called coordinate) divisors $(D)$ of distinct topology and we classify those surfaces with their possible deformations inside the pCICY 3-fold, which turn out to be satisfying $1 \leq h^{2,0}(D) \leq 7$. We also present a classification of the so-called ample divisors for all the favorable pCICYs which can be useful for fixing all the (saxionic) K\"ahler moduli through a single non-perturbative term in the superpotential. We argue that this relatively unexplored pCICY dataset equipped with the necessary model building ingredients, can be used for a systematic search of physical vacua. To illustrate this for model building in the context of type IIB CY orientifold compactifications, we present moduli stabilization with some preliminary analysis of searching possible vacua in simple models, as a template to be adopted for analyzing models with a larger number of K\"ahler moduli. 
}


\begin{document}

\maketitle

\bigskip

\section{Introduction}
\label{sec_intro}

To current date, String Theory is the most promising approach for the unification of quantum gravity with the other three fundamental interactions. In order to pursue such unification at a quantitative level, it is necessary to build explicit phenomenological models, which will reproduce known physics at the TeV scale, as well as predict new experimentally testable effects at higher energies. The construction of such models is precisely one of the core goals of the field of String Phenomenology. 

In this paper, we restrict our focus on type IIB model building. Any phenomenological model typically relies on an a priori choice of the geometry for the $6$-real dimensional compactification space, as well as discrete choices as for example the orientifold involution and flux quanta. Needless to say, different choices for these ingredients will inevitably lead to different 4d effective physics, which in turn has to be compared with current experiments. Therefore, in order to make sensible model building, it is of utmost importance to understand what are the possibilities for these choices.

Regarding the choice of the compactification space, this is typically assumed to be a Calabi-Yau (CY) manifold. Up to date, various explicit constructions for such manifolds exist. A very famous set is that of ``projective complete intersections Calabi-Yaus" (pCICYs) \cite{Candelas:1987kf}.\footnote{In fact, there exist other explicit and interesting constructions of CICY manifolds, such as the generalized CICY (gCICY) case of \cite{Anderson:2015iia} and the toric CICY (tCICY) as used in \cite{Cicoli:2021dhg}. For this reason, we call projective CICYs as pCICY 3-fold, just to distinguish them with the others.} These pCICY geometries are defined as the zero-locus $X$ of a set of polynomial equations in an ambient space $\mathcal{A}$ given by complex projective spaces. Another very famous construction for compact CY manifolds is that of Kreuzer and Skarke \cite{Kreuzer:2000xy}. Manifolds in this class far exceed the pCICYs in number, and are given by the zero locus of a section of the anticanonical bundle in a toric Fano 4-fold ambient space. We neglect these constructions in this paper. Given the vast number of possible compactification spaces, a natural question is how to select a specific one in order to build a phenomenological model. Some compactification spaces will indeed have properties that are somehow preferred, or more useful than others, in the model building.

This paper sheds some light on the previous question. In particular, we pursue a more systematic study of the phenomenological viability of the pCICYs dataset. Despite the fact that historically this was the first closed construction for Calabi-Yau manifolds, geometries in this dataset have remained only partially explored, so far, in the context of moduli stabilization and the subsequent phenomenology. In this regard, a systematic approach can be something well deserved. The reason for the choice to restrict our analysis to this construction only relies on the fact that pCICYs are particularly well-behaved and interesting for many reasons. First, the dataset is  closed and complete.\footnote{Here by \emph{complete} we mean that any other zero-locus of a set of polynomials in an ambient space of products of projective spaces which is \emph{not} in the list, has been proven to be real diffeomorphic to one already present in the pCICY list.} Secondly, for all but 70 of the pCICYs, it has been shown \cite{Anderson:2017aux} that all the divisors of $X$ are simply pullbacks of the divisors of $\mathcal{A}$. When this happens, we say that the pCICY is \emph{favorable}. In many cases, even the K\"ahler cone of $\mathcal{A}$ is isomorphic to that of $X$. These two properties simplify much of the analysis, as compared to CY manifolds in the Kreuzer-Skarke (KS) dataset.

For pCICYs, most of the topological information (such as triple intersection numbers, second Chern class, Euler characteristics, Mori cone, K\"ahler cone and even the Gopakumar-Vafa invariants \cite{Gopakumar:1998ii,Gopakumar:1998jq}), needed for model building at different stages have been computed in a series of work \cite{Anderson:2017aux,Hosono:1994ax,Carta:2021sms}: This has to be contrasted with the case of the KS dataset which has been explored in detail only up to $h^{1,1} \leq 6$ \cite{Altman:2014bfa,Altman:2021pyc} or partially for $h^{1,1} = 7$ as per recent claims in \cite{Gao:2021xbs}.

To take the analysis at a step further, in the present work we also compute the divisor topologies for all the pCICY 3-folds with special attention on the favorable ones and except for those which are of product type. This makes the collection of topological information more suitable and in a ready-to-be-used format for phenomenological model building. In this paper, we focus mainly on divisors of $X$ given by the pullbacks of the hyperplane classes $H_i$ of the various $\mathbb{P}^{n_i}$ in $\mathcal{A}$. We call such divisors \emph{coordinate divisors}. We will not discuss topologies of divisors given by linear combinations of the $\left. H_i \right|_X$. We leave comments on this interesting point to further study. 
The knowledge of the divisors' topology is truly essential in the context of global model building in IIB. In fact, one crucial point in this setup is that K\"ahler moduli can not be stabilized perturbatively by fluxes \cite{Cremmer:1983bf,Giddings:2001yu} (as opposed to IIA models) and therefore one needs a non-perturbatively generated superpotential $W_{\text{np}}$ for them \cite{Witten:1996bn,Balasubramanian:2005zx}. One standard way to achieve such effect consist in wrapping 7-branes, or Euclidean 3-branes, on specific 4-cycles of the compactification manifold. Crucially, the non-perturbative effect will be generated only if the divisor corresponding to the wrapped 4-cycle is rigid,\footnote{There are of course other ways to employ non-rigid divisors in order to produce a non-perturbatively generated superpotential, as for example turning on worldvolume flux on the 7-brane worldvolume \cite{Bianchi:2011qh,Bianchi:2012pn}, or employing freely acting involutions to construct a CY quotient in which some non-rigid cycles of the covering space will become rigid.} among with other conditions \cite{Witten:1996bn}. 

Remarkably, we find that, despite the large number of possible pCICYs, only $11$ different topologies of divisors are allowed. We report this in Table \ref{tab_divisor-topologies}. Crucially, \textit{none} of them is rigid. Furthermore, we find that at a given $h^{1,1}$ of the CY 3-fold, it appears to be easier implementing exchange involutions in the pCICY dataset as compared to the KS dataset. This is due to the larger frequency of finding what is called as ``non-trivially identical divisors" (NIDs) of the CY 3-folds \cite{Gao:2013pra} which are relevant for constructing their orientifolds. This latter fact will generically lead to the presence of odd-moduli in the 4d Effective Field Theory (EFT), something that has attracted lots of attention in recent years and can be further explored for application in axionic phenomenology \cite{Cicoli:2012vw,Cicoli:2013cha,Cicoli:2013zha,Cicoli:2013mpa,Hebecker:2018yxs,Cicoli:2021tzt,Cicoli:2021gss,Carta:2021uwv, Gao:2013pra, Gao:2013rra,Gao:2014uha}, specially for driving inflation and dark matter models. Also, motivated by the proposal of performing moduli stabilization using the so-called ample divisors \cite{Bobkov:2010rf} we have presented a classification of the ample divisors for all the favorable pCICYs, which we find to be always a non-rigid divisor. However, using the rigidification techniques via turning on background fluxes \cite{Bianchi:2011qh, Bianchi:2012pn, Louis:2012nb}, such divisors may help in generating a non-perturbative superpotential term.

By an interplay of both analytical reasoning and coding (we will extensively use the \texttt{cohomCalg} program of \cite{Blumenhagen:2010pv,Blumenhagen:2011xn}), we produce a dataset listing all the coordinate divisor topologies, for all favorable CICYs. The database we provide is available as ancillary. Being quite sizable, we believe that such database could be implemented for studying non-supersymmetric vacua using various contemporarily popular techniques such as Machine Learning, Genetic Algorithms, and Neural Network (see, e.g., \cite{He:2018jtw,Constantin:2021for,Abel:2021ddu,Abel:2021rrj,Larfors:2021pbb} and references therein).

This paper is organized as follows. In Section \ref{sec:pCICYs} we review basic facts about the geometry of complete intersection Calabi-Yau manifolds, and some of their relevant properties. In Section \ref{sec:pCICYs-divtop} we discuss the topology of their divisors, carefully explaining how we got to our classification result. Section \ref{sec_ample} is devoted to analyzing the ample divisors. In Section \ref{sec:pCICYh11-0} we comment on pCICYs with orientifold involutions that can lead to odd moduli, and their relevance for phenomenology. In Section \ref{sec:modustabrigdiv} we discuss aspects of moduli stabilization for pCICYs whose divisors can be rigidified, for example, by the addition of worldvolume fluxes. In Section \ref{sec:concl} we collect conclusions, and further possible research directions.

\section{A preliminary on pCICYs}
\label{sec:pCICYs}

In this section, we introduce the pCICYs 3-folds~\cite{Candelas:1987kf,Anderson:2017aux}. Such manifolds are defined as the zero-locus of a set of $k$ homogeneous polynomials $p_j(z)$ in an ambient space given by $\mathcal{A}=\prod_i \mathbb{P}^{n_i}$, constrained by
\begin{equation}
    \sum_{i}n_i-k=3\fstop
\end{equation}
The multi-degrees of the polynomial equations with respect to the coordinates of the ambient space factors are encoded in a configuration matrix
\begin{equation}
\left[
\begin{tabular}{c|cccc}
$\PP^{n_1}$ &   $q_1^1$ & $\cdots$  & $q_k^1 $ \\
$\PP^{n_2}$  &   $q_1^2$ & $\cdots$  &$ q_k^2$  \\
$\vdots$ &   $\vdots$ & $\ddots$ & $\vdots$  \\ 
$\PP^{n_s} $&   $q_1^s $& $\cdots$ & $q_k^s $
\end{tabular}
\right] \fstop
\label{eq:configuration}
\end{equation}
Requiring the zero-locus of the $p_j$ to be a CY manifold, the vanishing condition for the first Chern class imposes
\begin{equation} \label{eq:CYpolynomialcondgen}
n_i+1=\sum_{j=1}^{k} q_j^i\coma \quad \forall \, i=1,\ldots,s \fstop
\end{equation}
Let $X$ be a pCICY with ambient space $\mathcal{A}$. If $h^{1,1}(X)=h^{1,1}(\mathcal{A})$ we say that $X$ is a favorable pCICY. All pCICYs apart from 70 are favorable~\cite{Anderson:2017aux}. 

Using the configuration matrix in Eq. \eqref{eq:configuration}, it is possible to compute also the intersection polynomial of any favorable pCICY. Let us quickly review how. Let us denote by $D_i$ the divisor of $X$ descending from the hyperplane class of the corresponding $\PP^{n_i}$ in the ambient space $\mathcal{A}$, and let $\hat{D}_i$ its Poincar\'e dual form. The triple intersection number $\kappa_{ijk}$ is then defined by
\begin{equation}
\kappa_{ijk}=\int_X \hat{D}_i\wedge \hat{D}_j\wedge \hat{D}_k\fstop
\end{equation}
Such integral can be lifted to an integral over the ambient space, by
\begin{equation}
\kappa_{ijk}=\int_{\mathcal{A}} \hat{D}_i\wedge \hat{D}_j\wedge \hat{D}_k \bigwedge_{m=1}^k\left(\sum_{n=1}^s q_m^n \hat{D}_n\right)\fstop
\label{eq:tripleintamb}
\end{equation}

Being the ambient space a Cartesian product, such integral factorizes over an integral over the single $\mathbb{P}^{n_i}$ factor. As a consequence of this, the only time when the integral in \eqref{eq:tripleintamb} can be different from zero is when, for each factor in ambient space, in the integrand, it appears exactly its volume form. This is, for every one of the $s$ $\mathbb{P}^{n_i}$ in ambient space, there must appear in the integrand a wedge product of $n_i$ Poincar\'e dual forms $\hat{D}_i$. The triple intersection number correspond then to the overall coefficient multiplying the volume form of the ambient space, in \eqref{eq:tripleintamb}.

\section{Divisor topologies of favorable pCICYs}
\label{sec:pCICYs-divtop}

In this section, we present a classification for the divisor topologies of the pCICYs database \cite{Anderson:2017aux}. However, aiming for exploring their utilities for phenomenological purposes, we will consider only the favorable examples. Moreover, there are a total of 57885 coordinate divisors with non-identical GLSM charges in the given pCICY 3-fold, and we plan to compute and classify the topologies of all such divisors. The details of this counting with respect to the $h^{1,1}$ of the pCICYs are presented in Table \ref{tab_number-of-space-and-divisors}.
 
\begin{table}[!htp]
\centering
\renewcommand{\arraystretch}{1.3}
\begin{tabular}{|c|c||c|c|} 
\hline
$h^{1,1}$ & \# pCICYs  & \# fav. pCICYs & \# divisors of fav. pCICYs \\
 \hhline{|=|=#=|=|}
 1 & 5 & 5 & 5 \\
 2 & 36 & 36  & 72  \\
 3 & 155 & 155 & 465 \\
 4 & 425 & 425 & 1700 \\
 5 & 856 & 856 & 4280\\
 6 & 1257 & 1257 & 7542\\
 7 & 1463 & 1462 & 10234\\   
 8 & 1328 & 1325 & 10600\\ 
 9 & 1036 & 1032 & 9288\\ 
 10 & 648 & 643 & 6430\\ 
 11& 372 & 368 & 4048\\ 
 12& 161 & 155 & 1860\\ 
 13& 72 & 68 & 884\\ 
 14& 22 & 18 & 252\\ 
 15 & 16 & 15 & 225 \\ 
 16 & 1 & 0 & 0 \\ 
 17-18 & 0 & 0 & 0 \\ 
 19 & 15 & 0 & 0 \\ 
  \hline
 Total \# & 7868 & 7820 & 57885 \\ 
 \hline
\end{tabular}
\caption{Number of pCICYs and their divisors for $1\leq h^{1,1} \leq 15$.}
\label{tab_number-of-space-and-divisors}
\end{table}

\subsection{Computation of Hodge data of the divisors}
\label{sec:Hodgediv}

For a generic divisor ($D$) of the CY 3-fold $X$, there are only four independent Hodge numbers in the Hodge diamond, namely $h^{0,0}, h^{1,0}, h^{2,0}$ and $h^{1,1}$. Two of these can be computed from the Euler characteristics $\chi(D)$ and the Arithmetic genus $\chi_{_h}(D)$ which are given by the following useful formulae (e.g., see \cite{Blumenhagen:2008zz,Collinucci:2008sq, Bobkov:2010rf,Cicoli:2016xae}),
\begin{equation}
\begin{split}
\chi(D) &= 2 h^{0,0} - 4 h^{1,0} + 2 h^{2,0} + h^{1,1}= \int_{X} \left(\hat{D} \wedge \hat{D} \wedge \hat{D} + c_2(X) \wedge \hat{D} \right)\coma \\
 \chi_{_h}(D) &= h^{0,0} - h^{1,0} + h^{2,0} = \frac{1}{12} \int_{X}\left(2\, \hat{D} \wedge \hat{D} \wedge \hat{D} + c_2(X) \wedge \hat{D} \right)\fstop \end{split}
\label{eq:chi-chih}
\end{equation}
Here we denote the second Chern class of the CY 3-folds as $c_2(X)$ and, once again, $\hat{D}$ denotes the 2-forms dual to the divisor class. Thus, after knowing $\chi(D)$ and $\chi_{_h}(D)$ of a divisor using the second Chern class and the triple intersection numbers, one is practically left with computing only two out of the four Hodge numbers.

To begin with, we have used the \texttt{HodgeDiamond} module of the \texttt{cohomCalg} package \cite{Blumenhagen:2010pv,Blumenhagen:2011xn} to compute the divisor topologies for each of the so-called ``coordinate divisors" of all the favorable pCICYs with $1 \leq h^{1,1} \leq 4$ and partially for $h^{1,1} = 5$. However, we have subsequently observed that the program gets too slow for $h^{1,1} \geq 5$. Given that we effectively need to know only two Hodge numbers after having $\chi(D)$ and $\chi_{_h}(D)$, we subsequently used a different module \texttt{Lambda0CotangentBundle} of the \texttt{cohomCalg} package which computes the Hodge numbers $h^{0,0}, h^{1,0}$ and $h^{2,0}$ of a given divisor. Although this module is faster as compared to \texttt{HodgeDiamond}, it is still quite slow for larger $h^{1,1}$ of the pCICY 3-folds.

From our explicit computations of divisor topologies of pCICYs with $1\leq h^{1,1} \leq 6$ which correspond to scanning through around 14000 divisor topologies of more than 2500 CY spaces, we observe that each of the divisors always have the following Hodge numbers:
\begin{equation}
\label{eq:conjecture}
h^{0,0}(D) = 1 \coma h^{1,0}(D) = 0\fstop
\end{equation}
Given that we are working with the favorable CY 3-folds, it could be anticipated that divisors are smooth, in particular connected and hence $h^{0,0}(D) = 1$ is quite expected. Such observations have been made at multiple occasions in the context of Kreuzer-Skarke database as well \cite{Cicoli:2018tcq,Cicoli:2021dhg}. Based on these findings, we conjecture that Eq. \eqref{eq:conjecture} should be true for all the divisors of the favorable pCICY 3-folds.\footnote{However, this does not remain true for the non-favorable spaces, as we will discuss in Appendix \ref{app:non-favorable}.} Subsequently, using the conjecture \eqref{eq:conjecture} along with the triple intersection number and the second Chern class of the CY 3-fold, one can compute the remaining Hodge numbers of the divisors as
\begin{equation}
h^{2,0}(D) = \chi_{_h}(D) - 1\coma h^{1,1}(D) = \chi(D) - 2 \chi_{_h}(D)\fstop
\end{equation}
So, the generic divisor topology for the favorable pCICYs turns out to be of the following form
\begin{equation}
\label{eq:divisor-topology}
D\equiv
\begin{tabular}{ccccc}
    & & 1 & & \\
   & 0 & & 0 & \\
$(\chi_{_h}-1)$ \quad & & $(\chi - 2 \chi_{_h})$ \quad & & \quad $(\chi_{_h} -1)$ \\
   & 0 & & 0 & \\
    & & 1 & & \\
\end{tabular}\fstop
\end{equation}
Moreover, using the relation in Eq. \eqref{eq:chi-chih}, it is easy to find that divisors with vanishing self triple intersection $\int_{\rm CY} \hat{D} \wedge \hat{D} \wedge \hat{D} = 0$ satisfy the following relation,
\begin{equation}
\label{eq:Dcube=0}
h^{1,1}(D) = 10 h^{0,0}(D) - 8 h^{1,0}(D) +10 h^{2,0}(D)\fstop
\end{equation}
It is interesting to note that $K3$ satisfies such a relation. Since we have shown explicitly that $h^{1,0}(D)=0$ for all the CICYs up to $h^{1,1}=6$. We conjecture that this is true at least for favorable CICYs. 

\subsection{Scanning results and Classification}
\label{sec:scan-class}

There are a total of 57885 non-identical coordinate divisors for the 7820 favorable pCICY 3-folds, however we find that there are only 11 distinct topologies which arise from these pCICYs. They are classified in Table \ref{tab_divisor-topologies}.

\begin{table}[!htp]
\centering
\renewcommand{\arraystretch}{1.2}
\begin{tabular}{|Sc||Sc|Sc|Sc|Sc|Sc|} 
\hline
\shortstack{Sr.\\\#} & \shortstack{Divisor topology\\$\{h^{0,0}, h^{1,0}, h^{2,0}, h^{1,1}\}$} & \shortstack{frequency\\(57885 divisors)}  & \shortstack{frequency\\(7820 spaces)} & \shortstack{$h^{1,1}$\\(pCICY)} & \makecell{\vspace{0.2cm} $\displaystyle{\int_{_{\rm CY}} \hat{D}^3}$} \\
\hhline{|=#=|=|=|=|=|} 
T1 & $K3 \equiv \{1, 0, 1, 20\}$ & 30901 & 7736 & 2-15 & 0 \\
T2 & $\{1, 0, 2, 30\}$ & 22150 & 7436 & 2-15 & 0 \\
T3 & $\{1, 0, 3, 38\}$ & 3372 & 2955 & 2-13 & 2 \\
T4 & $\{1, 0, 3, 36\}$ & 91 & 91 & 3-13 & 4 \\
T5 & $\{1, 0, 4, 46\}$ & 714 & 690 & 2-11 & 4 \\
T6 & $\{1, 0, 4, 45\}$ & 283 & 277 & 1-11 & 5 \\
T7 & $\{1, 0, 4, 44\}$ & 91 & 91 & 2-11 & 6 \\
T8 & $\{1, 0, 5, 52\}$ & 198 & 198 & 1-9 & 8 \\
T9 & $\{1, 0, 5, 51\}$ & 28 & 28 &  1-9 & 9 \\
T10 & $\{1, 0, 6, 58\}$ & 42 & 42 & 1-7 & 12 \\
T11 & $\{1, 0, 7, 64\}$ & 15 & 15 & 1-5 & 16 \\
\hline
\end{tabular}
\caption{Divisor topologies for favorable pCICYs and their frequencies of appearance.}
\label{tab_divisor-topologies}
\end{table}

Based on the classification of the distinct topologies, we make the following interesting observations:
\begin{itemize}
\item We can see that the $K3$ surfaces and the so-called special deformation (SD) divisors\footnote{There has been a trend of denoting the divisors with $h^{2,0}(D) > 1$ as `special deformation' divisors  \cite{Gao:2013pra}. In this regard, the simplest one corresponds to the Hodge number $\{1, 0, 2, 30\}$ which for example also appears as the first three divisors of the famous $\mathbb{W}\CC\PP^4[11169]$ model used for realizing LVS.} which we denote as T2 in the collection of divisor topologies, appear in most of the pCICYs while as the number of deformation $h^{2,0}(D)$ increases, the corresponding frequencies of occurrence is reduced. 

\item We observe that the self triple intersection number $\int_{_{\rm CY}} \hat{D} \wedge {\hat D} \wedge {\hat D}$ vanishes for divisor topologies T1 and T2 which satisfy $h^{1,1}(D) = 10 + 10 \, h^{2,0}(D)$. This is well anticipated to hold for $h^{0,0}(D) =1$ and $h^{1,0}(D) = 0$ using Eq. \eqref{eq:Dcube=0}.

\item Moreover, for other divisors of given $h^{2,0}(D)$ and having multiple topologies, we observe that the frequency of occurring those topologies is more for larger $h^{1,1}(D)$. So it appears that divisor topologies tend to be more frequent towards $\int_{_{\rm CY}} \hat{D} \wedge {\hat D} \wedge {\hat D}$ tending to zero values.

\item We observe that there are no rigid surfaces present in the divisor topologies. This could naively be anticipated because these pCICYs are defined by multi-hypersurface constraints in the product of $ {\mathbb P}^n$'s and after considering the favorable cases only, all the divisors of the CY are descending from the divisors of the ambient space. Thus, for the least possible scenario where $ {\mathbb P}^1$ appears in the product of $ {\mathbb P}^n$'s, there would be at least two coordinates (corresponding to the $ {\mathbb P}^1$) with same GLSM charges, which can be subsequently deformed into each other.

\item We also observe that there are no ${\mathbb T}^4$ surfaces present as a coordinate divisor of the favorable pCICYs.

\item It is an interesting observation to make that the topology T11 appears $n$ times, where $1\leq n \leq 5$ respectively for $1\leq h^{1,1}($pCICY$) \leq 5$ summing up to a total of 15, i.e., it appears once for $h^{1,1}($pCICY$)=1$, twice for $h^{1,1}($pCICY$) = 2$, three times for $h^{1,1}($pCICY$)=3$, four times for $h^{1,1}($pCICY$)=4$ and five times for $h^{1,1}($pCICY$) = 5$. It does not appear for larger values of $h^{1,1}($pCICY$)$ at all.

\end{itemize}

\section{Finding the ample divisor}
\label{sec_ample}

It has been proposed in \cite{Bobkov:2010rf} that using an appropriate form of an ample divisor which is rigid as well, one can have a superpotential contribution of the following form, 
\begin{equation}
\label{eq:Wnp-ample}
W_{\text{np}} = A \, \exp\left[-i\sum_{\alpha = 1}^{h^{1,1}_+(CY)} a_\alpha T_\alpha\right]\coma
\end{equation}
where $T_\alpha$ are the K\"ahler moduli. These are interesting because a single superpotential term can stabilize all the saxions of the $T_\alpha$ moduli while fixing one overall combination of the $C_4$ axions, leaving $h^{1,1}_+(CY) -1$ of them still unfixed. 

\subsection{Criteria for ampleness}

There can be multiple different ways of defining an ample divisor for a given smooth algebraic variety $X$. One criterion is the so-called Nakai-Moishezon criterion for ampleness in which a divisor $D_A$ of X is ample if the following holds,
\begin{equation}
\label{eq:Nakai-Moishezon}
D_A^{\rm dim(Y)}\cdot Y > 0\coma
\end{equation}
for every closed subvariety of $Y$ of $X$. In the language of intersection theory, this criterion results in the following conditions which are necessary for $D_A$ to being ample,
\begin{equation}
\label{eq:ample-1}
    \begin{split}
        \int_{CY} D_A \wedge D_A \wedge D_A &> 0\coma\\
        \int_{CY} D_A \wedge D_A \wedge J_i &> 0 \quad \forall \, \, i\coma\\ 
        \int_{CY} D_A \wedge J_i \wedge J_j &> 0 \quad \forall \, \, i \neq j\fstop
    \end{split}
\end{equation}
An ample divisor of a CY 3-fold is also described as a codimension 1 surface with ample canonical bundle, which is also called as surface of {\it general type} described by the following two Chern numbers \cite{Bobkov:2010rf},
\begin{equation}
\label{eq:c1-c2-positive}
c_1^2 = \int_D c_1(D)^2 > 0\coma c_2 = \int_D c_2(D) >0\fstop
\end{equation}
Let us note that \cref{eq:c1-c2-positive} regarding the positivity of $c_1$ and $c_2$ can also be expressed in terms of Euler characteristic $\chi(D)$ and the Arithmetic genus $\chi_{_h}(D)$ of a given divisor. It turns out that using \cref{eq:chi-chih} we have the following equivalent condition,
\begin{equation}
\label{eq:ample-2}
6 \, \chi_{_h}(D) < \chi(D) < 12\, \chi_{_h}(D)\fstop
\end{equation}
In addition, there is another condition known as Bogomolov-Miyaoka-Yau inequality which holds for surfaces of {\it general type}, which is given as,
\begin{equation}
\label{eq:BMYineq}
c_1^2 \leq 3 \, c_2 \quad \Leftrightarrow \quad \chi(D) \geq 3 \chi_{_h}(D)\fstop
\end{equation}
However, this condition is weaker than the inequalities of \cref{eq:ample-2}. Let us note that the inequalities in \eqref{eq:ample-2} do not involve the boundaries, and hence any surfaces which satisfy this as a limiting case will not be of general type. For example, a $K3$ surface satisfies $\chi(K3) = 24 = 12 \, \chi_{_h}(K3)$ and for the torus ${\mathbb T}^4$ we have $\chi({\mathbb T}^4)  = 0 = \chi_{_h}({\mathbb T}^4)$ and therefore both of these topologies are not of {\it general type}. In addition, note that the second topology T2 in our collection in Table \ref{tab_divisor-topologies} satisfies $\chi({\rm T2}) = 36 = 12 \, \chi_{_h}({\rm T2})$ and hence this does not satisfy the condition \eqref{eq:ample-2}.

In \cite{Bobkov:2010rf}, a couple of favorable properties of ample divisors have been presented which can be useful as a cross-check while classifying the ample divisors. For example, the Kodaira vanishing condition,
\begin{equation}
h^q(X, {\cal O}(D) ) = 0\coma q > 0\coma
\end{equation}
and the Lefshetz hyperplane theorem \cite{Hubsch:1992nu} implying that,
\begin{equation}
\begin{array}{rclcrcl}
    h^{p,q}(X) &=& h^{p,q}(D) &\text{\hspace{0.5cm}} &\forall \,\, p+q &<& 2\coma\\
    \pi_i(X) &=& \pi_i(D) & &\forall \,\,i & \leq &1\fstop
\end{array}
\end{equation}
Using these necessary conditions one finds that an ample divisor $D_A$ has to satisfy the following conditions,
\begin{equation}
\label{eq:ample-0}
h^{0,0}(D_A) = 1\coma h^{1,0}(D_A) = 0\fstop
\end{equation}
Notice that \cref{eq:ample-0} presents the same requirement as what was conjectured in \cref{eq:conjecture} to hold for all the divisors of the pCICYs, which hints that there might be an ample amount of abundance of these so-called ample divisors in the pCICY database.

As an alternative definition presented in \cite{Bobkov:2010rf}, a divisor is ample {\it iff} the Poincar\'e isomorphism $H_4(X, {\mathbb Z}) \simeq H^2(X, {\mathbb Z})$ identifies it with a cohomology class that can be represented by the K\"ahler form of a smooth CY metric. Given that the smooth CY metric defines a class $[\omega] \in K_c(X) \subset H^{1,1}(X)$, where $K_c(X)$ denotes the K\"ahler cone of the 3-fold $X$. Using the K\"ahler cone generators $J_i$ as the basis of the $(1,1)$-forms, an ample divisor may be represented as,
\begin{equation}
\label{eq:ample-sumDi}
D_A = \sum_{i = 1}^{h^{1,1}_+} n_i \, J_i\coma \forall \, \, n_i > 0\fstop
\end{equation}
This definition appears to play a direct r\^ole in understanding the ample divisors of the pCICYs in the database as we will discuss later on. 

\subsection{Criteria for rigidity}

In the context of type IIB compactification on CY $3$-folds, the holomorphic superpotential can receive non-perturbative contributions from $E3$-instantons or gaugino condensation effects on $D7$-brane wrapping appropriate (rigid) divisors (D). In fact, a necessary condition for the non-perturbative superpotential contribution, in the absence of any fluxes, can be given as \cite{Witten:1996bn}
\begin{equation}
\label{eq:unit-chih-1}
\chi_{_h}(D) = \sum_{i = 0}^2 {(-1)}^{i} \, h^{i,0}(D) = 1.
\end{equation}
For a given divisor (D) wrapping an $E3$-instanton or a $D7$-brane with fluxes, this is known as Witten's ``unit Arithmetic genus condition". However, in the presence of fluxes, the topological condition (\ref{eq:unit-chih-1}) can be relaxed and it is possible that divisors with $\chi_{_h}(D) > 1$ can also contribute to the non-perturbative superpotential \cite{Gorlich:2004qm,Kallosh:2005yu,Tripathy:2005hv,Saulina:2005ve,Kallosh:2005gs,Bergshoeff:2005yp}.

For this purpose, a class of divisors known as the ``rigid" divisors are of particular interest. A divisor $(D)$ is defined to be rigid if it cannot be deformed inside the CY 3-fold, which corresponds to have $h^{2,0}(D) = 0$. One can think of a divisor as zeros of a section in a line bundle ${\cal O}(D)$ and subsequently the rigidity is correlated with the section being unique up to an overall constant, i.e. $h^0(X, {\cal O}(D)) = 1$, e.g. see \cite{Bobkov:2010rf}. Therefore, for a rigid smooth divisor one generically has:
\begin{equation}
\chi_{_h}(D) = 1 - h^{1,0}(D)\coma \chi(D) = 2 - 4 h^{1,0} + h^{1,1}(D).
\end{equation}
However, for our case of coordinate divisors of the favorable pCICY $3$-folds being smooth and simply connected, the above conditions following from the rigidity of the divisor (D) simplify to the following form,
\begin{equation}
\chi_{_h}(D) = 1\coma \chi(D) = 2 + h^{1,1}(D)\fstop
\end{equation}
Note that, for our case of coordinate divisors we always have $\chi_{_h}(D) = 1 + h^{2,0}(D) > 1$, and so we do not have any rigid divisors as can be seen from the classification presented in Table \ref{tab_divisor-topologies}. Failing to satisfy the necessary condition given in Eq.~(\ref{eq:unit-chih-1}), a priory, this observation may appear as an obstruction to have non-perturbative effects in models based on pCICY dataset. However this is not the case as we further elaborate now.

For model building in the framework of type IIB orientifold compactifications with $O3/O7$-planes, in order for an $E3$-instanton to contribute to the superpotential, it has to be an ${\cal O}(1)$ instanton wrapping a transversely invariant divisor of the CY $3$-fold as follows from a detailed zero-mode analysis, e.g. see \cite{Blumenhagen:2010ja}. In this regard, one  defines another index $\chi_{_h}^\sigma (D)$ for taking care of the splitting into the even/odd sector under the orientifold involution $\sigma$ in the following manner \cite{Blumenhagen:2010ja, Bianchi:2011qh},
\bea
\label{eq:split-chih}
& & \chi_{_h}^\sigma (D)  = \sum_{i = 0}^2 {(-1)}^{i} \, \left(h^{i,0}_{+}(D) - h^{i,0}_{-}(D) \right) = \chi_{_h}^{+}(D) - \chi_{_h}^{-}(D),
\eea
where the last equality defines the even/odd quantities $\chi_{_h}^{\pm}(D)$. The reason for defining this quantity $\chi_{_h}^\sigma (D)$ can be understood from the fact that it can be also expressed as the arithmetic genus of a divisor corresponding to the CY $4$-fold,
\bea
\label{eq:split-chih-1}
& & \chi_{_h}^\sigma (D) = \sum_{i = 0}^3 {(-1)}^{i} {\frak h}^{i}(D),
\eea
where
\bea
{\frak h}^{i}(D) = \{h^{0,0}_{+}(D),  h^{0,0}_{-}(D) + h^{1,0}_{+}(D), h^{1,0}_{-}(D) +h^{2,0}_{+}(D), h^{2,0}_{-}(D)\}.
\eea
Therefore, it can be interpreted as zero-mode counting of the $M5$-brane instantons in the $M/F$ theory language, e.g. see Table 4 of \cite{Blumenhagen:2010ja}, and subsequently leads to the following necessary condition for non-perturbative superpotential contribution,
\begin{equation}
\label{eq:unit-chih-2}
\chi_{_h}^{+}(D) - \chi_{_h}^{-}(D) = 1\fstop
\end{equation}
For rigid smooth divisors of a CY 3-fold with $h^{1,0}(D) = 0 = h^{2,0}(D)$, the necessary condition in Eq.~\eqref{eq:unit-chih-1} is automatically satisfied. Moreover, in this case the transverse invariance of the smooth divisor to have an $O(1)$ instanton demands that $h^{0,0}(D) = h^{0,0}_+(D) = 1$ and hence $\chi_{_h}^{+}(D) = 1$ and $\chi_{_h}^{-}(D) = 0$ for usual standard reflection involutions and hence it satisfies the necessary condition in Eq.~(\ref{eq:unit-chih-2}) as well. However for non-rigid divisors with $\chi_{_h}(D) > 1$ which corresponds to our case of coordinate divisors of pCICYs, one can find some appropriate involutions along with a set of discrete symmetries such that $\chi_{_h}^{+}(D) - \chi_{_h}^{-}(D) = 1$, and hence such divisors can also contribute to the superpotential despite having $\chi_{_h}(D) > 1$; for explicit CICY examples, see \cite{Bobkov:2010rf}. 

Moreover, there are additional prescriptions to ``rigidify" a non-rigid divisor by turning on the magnetic fluxes which helps in lifting the extra unwanted zero modes to facilitate non-perturbative superpotential contribution. For example, the first three coordinate divisors of the well known swiss-cheese CY ``11169" (defined as a degree-18 hypersurface in WCP$^4[1,1,1,6,9]$) correspond to the topology T2 in our classification given in Table \ref{tab_divisor-topologies}, and therefore have $h^{2,0}(D) = 2$. Such a divisor has been rigidified in \cite{Louis:2012nb}. Similarly, the first four coordinate divisors of the CY ``11114" (defined as a degree-8 hypersurface in WCP$^4[1,1,1,1,4]$) have $h^{2,0}(D) = 3$, and these have been rigidified in   \cite{Bianchi:2011qh,Bianchi:2012pn}. Similarly we argue that the coordinate divisors of the pCICYs can also be rigidified to guarantee the non-perturbative superpotential contributions for K\"ahler moduli stabilization.

\subsection{Criteria for rigid ample divisors}

Let us note that demanding a divisor to be rigid as well as ample turns out to be quite restrictive. The criterion \eqref{eq:c1-c2-positive} for ampleness along with the Arithmetic genus condition \eqref{eq:unit-chih-1} results in the following condition,
\begin{equation}
\label{eq:rigid-ample-1}
6 < \chi(D) < 12\coma
\end{equation}
which leaves only five possibilities for the Euler characteristic of the divisor. Moreover, it was argued in \cite{Bobkov:2010rf} that surfaces arising as ample divisors on a CY 3-folds have $|\pi_1(D)| = |\pi_1(X)| < \infty$ which results in the following equivalent constraints,
\begin{equation}
\label{eq:rigid-ample-2}
1 \leq c_1^2 \leq 5 \quad \Leftrightarrow \quad 1 \leq 12\, \chi_{_h}(D) - \chi(D) \leq 5\fstop
\end{equation}
Given that $\chi(D)$ and $\chi_{_h}(D)$ can only take integral values, one can observe that the condition in \cref{eq:rigid-ample-2} is already contained in \cref{eq:rigid-ample-1} which follows from \cref{eq:ample-2}. Subsequently, we have the following five possibilities,
\begin{equation}
\label{eq:rigid-ample-3}
    \begin{split}
\{\chi_{_h}, \, \chi \} &: \quad \{1, 7\}\coma  \{1, 8\}\coma  \{1, 9\}\coma  \{1, 10\}\coma  \{1, 11\}\,;\\
\left\{c_1^2, \, c_2 \right\} &: \quad \{5, 7\}\coma  \{4, 8\}\coma  \{3, 9\}\coma  \{2, 10\}\coma  \{1, 11\}\coma
    \end{split}
\end{equation}
where in order to have some comparison, we have presented the respective values for the $\left(c_1^2, c_2\right)$ pair corresponding to their $(\chi_{_h}, \chi)$ values in the second row. A surface with $\chi_{_h} = 1$ and $\chi = 10$ known as Campedelli surface was mentioned in \cite{Bobkov:2010rf}. Note that this surface have the same Hodge diamond as of a del Pezzo surface dP$_7$, however both are entirely different, given that for del Pezzo surfaces the anti-canonical bundle is ample.

\subsection{Classification}

In our detailed analysis we have realized that there are only 11 types of topologies for the divisors of the favorable pCICY 3-folds as mentioned in Table \ref{tab_divisor-topologies}. We find that all the divisors satisfy the necessary condition \eqref{eq:ample-0} for being ample, however only 9 out of 11 satisfy the additional condition of ampleness given in \cref{eq:ample-2}. It turns out that the topologies T1 and T2 which constitute more than 90\% of the total divisor topologies we have analyzed, do not satisfy \eqref{eq:ample-2} as we have mentioned earlier. Also, given that we do not have rigid divisors in the list of 11 topologies, we do not have any divisor which is ample as well as rigid at the same time. However, there are methods of rigidifying the divisors \cite{Bianchi:2011qh,Bianchi:2012pn} which can serve for the purpose of generating non-perturbative superpotentials, and so we still seek for those divisors which could be ample and subsequently rigidified by any of the mechanism proposed in \cite{Bianchi:2011qh,Bianchi:2012pn}. The main benefit of having such a divisor is to fix all the saxionic moduli using a single exponentially suppressed term in the superpotential.

For constructing an ample divisor of the kind \eqref{eq:ample-sumDi} we consider a combination with $n_i = 1$ for all $i \in h^{1,1}_+$(CY), i.e. we propose the following form for an ample divisor $D_A$ and then we will check all the conditions which we have listed by computing the divisor topologies of $D_A$,
\begin{equation}
\label{eq:ample-sum}
D_A = \sum_{i =1}^{h^{1,1}_+{\rm (CY)}} J_i\fstop
\end{equation}
We note that for $h^{1,1}_+(CY) = 1$, all the five examples do satisfy the criterion \eqref{eq:ample-sum} in the sense that each of their divisors satisfy the ampleness requirement. Using \texttt{cohomCalg} \cite{Blumenhagen:2010pv,Blumenhagen:2011xn} we explicitly checked that the requirements in \cref{eq:ample-0} hold for $D_A$ of the pCICYs with $h^{1,1}_+$(CY) $ < 7$. Subsequently, we conjecture that it should hold for $h^{1,1}_+$(CY) $\geq 7$ as well. To corroborate our argument, we also mention a case of non-favorable example studied in \cite{Bobkov:2010rf} using a CICY 3-fold with $h^{1,1}$(CY) $= 19$ and resulting in an ample divisor of the type we propose in \cref{eq:ample-sum}. For each of the favorable pCICY 3-fold, we checked that the proposed divisor in \cref{eq:ample-sum} satisfy the constraints in \cref{eq:ample-0,eq:ample-1,eq:ample-2}. So the only task remained for guaranteeing them to contribute to the superpotential is to rigidify the $D_A$ for each of the pCICY 3-fold following the prescriptions in \cite{Bianchi:2011qh,Bianchi:2012pn}. As the underlying mechanisms used in these prescriptions are very much dependent on the detailed specific such as choice of involution, brane setting and other ingredients of the local model building, it requires a separate comprehensive attention which we plan for a future work. For illustrating moduli stabilization in the current work, we will simply assume that such an ample divisor can always be rigidified (choosing appropriate fluxes) and hence can be used for the generation of non-perturbative superpotential term of the type mentioned in \cref{eq:Wnp-ample}.


\section{pCICY orientifolds with $h^{1,1}_- \ne 0$}
\label{sec:pCICYh11-0}

Given that there are only 11 distinct topologies for a huge number of coordinate divisors of the pCICY 3-folds, one can expect that there is a significant possibility of finding what is called as ``non-trivially identical divisors" \cite{Gao:2013pra}. These are divisors which have the same Hodge diamonds but different GLSM charges. In that sense, they are non-trivially identical inside the CY 3-fold and can help in constructing the even/odd pairs of divisor classes under the orientifold involution by exploiting some exchange symmetries. Explicit models with $h^{1,1}_- \neq 0$ have been studied in \cite{Lust:2006zg,Lust:2006zh,Blumenhagen:2008zz,Cicoli:2012vw,Gao:2013rra,Gao:2013pra,Gao:2014uha,Carta:2020ohw,Carta:2021sms, Carta:2021uwv,Cicoli:2021tzt,Altman:2021pyc,Cicoli:2021gss}.

We argue that having a significantly large number of type IIB O3/O7 models with odd axions is relatively easier in the pCICYs dataset \cite{Anderson:2017aux} as compared to the KS dataset \cite{Kreuzer:2000xy}. In fact, it has been recently found in \cite{Altman:2021pyc} that the total number of polytopes of KS database which could result in consistent orientifolds using exchange involution is ${\cal O}(1400)$ for a total number of polytopes ${\cal O}(23000)$ corresponding to $h^{1,1}({\rm CY}) \leq 6$. Also, it has been further predicted that the percentage of such consistent polytopes reduces with increasing $h^{1,1}({\rm CY})$ \cite{Gao:2021xbs}. On the other hand, for the pCICY dataset, we find that only 0.014\% spaces are a priory ruled out due to the absence of NIDs. 

To illustrate these arguments, we will present some explicit orientifold examples with large $h^{1,1}_- $. This can be useful for several axion phenomenology purposes, e.g., in building multi-axion models of inflation or dark matter.

\subsection{Classification of models}
\label{sec:clasmodel}

If such a pair of divisors exist, say $D_1$ and $D_2$, which are of NID type and correspond to the coordinates $x_1$ and $x_2$ of the CY 3-fold, then one can consider a combination of the two divisor classes,
\begin{equation}
D_\pm = D_1 \pm D_2\coma
\end{equation}
such that $D_+$ is even and $D_-$ is odd under the exchange involution $x_1 \leftrightarrow x_2$. This subsequently leads to a non-trivial $(1,1)$-cohomology for the odd sector, i.e. $h^{1,1}_- \neq 0$. It is relatively easier to classify such spaces in the context of pCICYs as opposed to the CY 3-folds of the KS database. The reason being the fact that the respective ambient spaces of the pCICY 3-folds are expressed as products of ${\mathbb P}^m$'s for some $m \geq 1$ and therefore it is more likely that if the topologies of the two divisors corresponding to two different ${\mathbb P}^m$'s matches, then their SR ideals have good chance to respect that exchange involution, i.e. $x_1 \leftrightarrow x_2$. Such a situation does not arise so naturally in the KS database examples, as sometimes SR ideals do not respect this exchange symmetry under which two NIDs are exchanged. Analyzing the divisor topologies of all the favorable pCICYs we find that there are only 111 (out of 7820) spaces which do not possess such a pair of NIDs and hence are not suitable for having  a non-trivial (1,1)-cohomology for the odd sector. The remaining space, in principle, can have exchange involutions such that $h^{1,1}_- \neq 0$. A more specific classification about when a pCICY does not admit NIDs is given in the following Table \ref{tab_ZeroNIDs}.

\begin{table}[!htp]
\centering
\renewcommand{\arraystretch}{1.2}
\begin{tabular}{|c||c|c|c||c|c|} 
\hline
$h^{1,1}({\rm pCICY})$ & 1 & 2  & 3 & 4 & 5-15 \\
\hhline{|=#=|=|=#=|=|}
Total \# of fav. pCICYs & 5 & 36 & 155 & 425 & 7199 \\
\# of pCICYs without NIDs & 5 & 31 & 67 & 8 & 0 \\
\# of pCICYs with $h^{1,1}_- \neq 0$ & 0 & 5 & 88 & 417 & 7199 \\
\hline
\end{tabular}
\caption{Number of favorable pCICYs without any NIDs which can always result in $h^{1,1}_- = 0$.}
\label{tab_ZeroNIDs}
\end{table}

\noindent
Thus, given the richness of the pCICY dataset for the NIDs, it would be good for pursuing type IIB model building with odd-moduli which has attracted a significant amount of interest in recent years \cite{Gao:2014uha,Cicoli:2021tzt}. Let us consider some explicit examples for demonstrating these findings. 

\subsection{Some explicit constructions}
\label{sec:explcontr}

In this section, we present some explicit construction to illustrate the recipe of exchanging the NIDs to get $h^{1,1}_- \neq 0$. We also provide relevant topological data for some orientifolds which can be subsequently used for any phenomenological purpose on the lines of \cite{Lust:2006zg,Lust:2006zh,Blumenhagen:2008zz,Cicoli:2012vw,Gao:2013rra,Gao:2014uha,Carta:2020ohw,Carta:2021sms, Carta:2021uwv,Cicoli:2021tzt,Cicoli:2021gss}.

\subsubsection{$h^{1,1}_-(\rm pCICY) = 1$}

To start with, let us consider the minimal orientifold construction with non-trivial $h^{1,1}_-$ which corresponds to having $h^{1,1}(\rm pCICY) = 1_+ + 1_-$. For $h^{1,1} = 2$ case there are only five pCICY examples which can result in NIDs, and those are given in Table \ref{tab_NIDs-h11=2}.
\begin{table}[!htp]
\centering
\renewcommand{\arraystretch}{1.2}
\begin{tabular}{|Sc||Sc|Sc||Sc|Sc|Sc|} 
\hline
\shortstack{Sr.\\\#} & \shortstack{Space \#\\$\{7820,7890\}$} & \makecell[c]{\vspace{0.3cm}$\{h^{1,1}, h^{2,1}\}$}  & \shortstack{NIDs\\Topologies} & \makecell[c]{\vspace{0.3cm}$\kappa_{ijk}$}  & \makecell[c]{\vspace{0.1cm}$\displaystyle{\int_{_{\rm CY}} c_2 \wedge J_i}$} \\
\hhline{|=#=|=#=|=|=|}
1 & $\{7574, 7644\}$ & $\{2, 46\}$ & \{1, 0, 4, 46\} & $\{4, 12, 12, 4\}$ & $\{52, 52\}$ \\
2 & $\{7691, 7761\}$ & $\{2, 52\}$ & \{1, 0, 4, 45\} & $\{5, 10, 10, 5\}$ & $\{50, 50\}$ \\
3 & $\{7729, 7799\}$ & $\{2, 55\}$ & \{1, 0, 3, 38\} & $\{2, 7, 7, 2\}$ & $\{44, 44\}$ \\
4 & $\{7793, 7863\}$ & $\{2, 66\}$ & \{1, 0, 3, 38\} & $\{2, 6, 6, 2\}$ & $\{44, 44\}$ \\
5 & $\{7814, 7884\}$ & $\{2, 83\}$ & \{1, 0, 2, 30\} & $\{0, 3, 3, 0\}$ & $\{36, 36\}$ \\
\hline
\end{tabular}
\caption{Number of favorable pCICYs for $h^{1,1} = 2$ and having NIDs to give $h^{1,1}_- = 1$. Here, the divisor topologies are represented with the Hodge numbers $\{h^{0,0}, h^{1,0}, h^{2,0}, h^{1,1}\}$. Also, the intersection section numbers $\kappa_{ijk}$ are collected as $\{\kappa_{111}, \kappa_{112}, \kappa_{122}, \kappa_{222}\}$.}
\label{tab_NIDs-h11=2}
\end{table}

\noindent
Note that the last CY of Table \ref{tab_NIDs-h11=2} represents a bi-cubic in ${\mathbb P}^2 \times {\mathbb P}^2$, and it also appears in the KS database. In fact, this is the only example of the KS database with $h^{1,1} =2$ which can split as $h^{1,1}_\pm({\rm CY}/{\cal O}) =1$ under exchange involution \cite{Gao:2013pra}. In addition to that, we observe that the pCICYs collection has four more of such examples. Let us also mention that such CY 3-folds can be useful for realizing the so-called exponentially flat flux vacua as recently reported in \cite{Carta:2021kpk} using the fifth CY in Table \ref{tab_NIDs-h11=2}.

One can observe that the intersection polynomials are symmetric under the exchange of $1 \leftrightarrow 2$ and hence the exchange symmetry exists. The even/odd divisors are constructed as a combination of these NIDs in the following manner,
\begin{equation}
J_\pm = J_1 \pm J_2\coma
\end{equation}
which leads to the following intersection polynomials for the five examples,
\begin{equation}
\begin{split}
I_3^\sigma & =  80J_+^3 -16 J_+ J_-^2\coma \\ 
I_3^\sigma & =  70 J_+^3 -10 J_+ J_-^2 \coma\\ 
I_3^\sigma & =  46J_+^3 -10 J_+ J_-^2\coma  \\
I_3^\sigma & =  40J_+^3 -8 J_+ J_-^2\coma \\
I_3^\sigma & =  18J_+^3 -6 J_+ J_-^2   \fstop 
\end{split}
\end{equation}
Notice that after orientifolding, the intersection numbers of the type $J_+^2 J_-$ and $J_-^3$ trivially vanish. Moreover, $J_+ J_-^2$ is always negative while $J_+^3$ is positive, which contribute to the geometric volume of the orientifolded CY 3-fold.


\subsubsection{$h^{1,1}_-(\rm pCICY) = 2$}

To illustrate such a splitting of the even/odd sector for a pCICY with $h^{1,1} = 4$ we present the model numbered as 7862 in the list of 7890 while it is numbered as 7792 in the collection of 7820 favorable pCICYs. The topological data for this CY 3-fold is given as
\begin{equation}
 \begin{array}{c}
    \renewcommand*{\arraystretch}{1.25}
\displaystyle{   \{h^{1,1}, h^{2,1}\} = \{4, 68\}\coma{\rm Topo}(J_i) \equiv \{1, 0, 1, 20\}\coma \int_{_{\rm CY}} c_2 \wedge J_i = 24\coma \forall i=1,\ldots,4}\\
\displaystyle{   I_3 = 2 J_1 J_2 J_3 + 2 J_1 J_2 J_4 + 2 J_1 J_3 J_4+ 2 J_2 J_3 J_4\coma}
    \end{array}
\end{equation}
which show that all the divisors in the basis are $K3$ surfaces, and the CICY is a $K3$-fibered CY 3-fold. Now there are multiple ways of choosing the even/odd combinations of NIDs, and one such possibility is considered as below,
\begin{equation}
J_{p_1} = J_1 + J_2\coma J_{p_2} = J_3 + J_4\coma J_{m_1} = J_1 - J_2\coma J_{m_2} = J_3 - J_4\fstop
\end{equation}
This leads to the following intersection polynomials after the orientifolding through the exchange involution $\{1 \leftrightarrow 2, 3 \leftrightarrow 4 \}$,
\begin{equation}
I_3^\sigma = 8 J_{p_1}^2 J_{p_2}  + 8 J_{p_1} J_{p_2}^2  - 8  J_{p_1} J_{m_2}^2  - 8  J_{p_2} J_{m_1}^2\fstop
\end{equation}
Similar to the previous case, we observe that the intersection numbers of the type $J_+^2 J_-$ and $J_-^3$ trivially vanish while $J_+ J_-^2$ are non-positive. 

\subsubsection{$h^{1,1}_-(\rm pCICY) = 7$}

To illustrate that having a significantly large number of odd axions is relatively easier with pCICYs,\footnote{Analyzing around 23000 reflexive polytopes leading to more than 500000 triangulations and nearly 100000 distinct CY geometries with $h^{1,1} \leq 6$ of the KS database \cite{Kreuzer:2000xy}, it turns out that one can only have $h^{1,1}_- \leq 3$ for the dataset in \cite{Altman:2021pyc}. However, this upper limit of $h^{1,1}_-$ being 3 could be anticipated, given that $h^{1,1}(CY) \leq 6$ for that dataset, and we need pairs of NIDs for constructing odd divisor $D_-$ in the basis.} now we present an explicit example with $h^{1,1}_- = 7$. For this purpose we consider the very first example of the pCICY dataset which has $\{h^{1,1}, h^{2,1}\} = \{15, 15\}$. The other topological data for this CY 3-fold is given as under,
\begin{equation}
\renewcommand*{\arraystretch}{2}
 \begin{array}{rlllllrll}
{\rm Topo}(J_i) & \equiv & \{1, 0, 1, 20\} \coma & \displaystyle{\int_{_{\rm CY}} c_2 \wedge J_i} & = & 24 \coma & i & \in & {\cal A} = \{2, 4, 7, 9, 10, 11, 12, 13, 15 \}\coma\\
{\rm Topo}(J_j) & \equiv & \{1, 0, 2, 30\}\coma & \displaystyle{\int_{_{\rm CY}} c_2 \wedge J_j} & = & 36\coma & j & \in & {\cal B} = \{1, 3, 5, 6, 8, 14 \}\fstop
    \end{array}
\end{equation}
The intersection polynomial in this example can be anticipated to be quite complicated given that there are 15  elements in the divisor basis. However, notice that the two divisor topologies for this example are T1 and T2 in our classification. Also, we recall that $\int_{CY} \hat{D}^3 = 0$ for both of these topologies as follows from \cref{eq:Dcube=0}. Moreover, for $K3$-fibered CY 3-fold, a theorem of Oguiso \cite{OGUISO:1993} (see also \cite{Schulz:2004tt}), tells that the fibered $K3$ divisor can only appear linearly in the intersection polynomial. So it is not only the cubics which vanish for the $K3$ divisor but also the quadratics. Therefore, the intersection polynomial is expected to have the following vanishing intersections,

\begin{equation}
\label{eq:intpoly-h11=15-1}
\int_{CY} J_i^3 = 0\coma \int_{CY} J_i^2 \, J_j = 0\coma \int_{CY} J_j^3 = 0 \coma \forall i \in {\cal A} \, \, {\rm and} \, \, j \in {\cal B}, 
\end{equation}
which we have verified by looking into the explicit form of the intersection polynomial. In fact, for $h^{1,1} = 15$, there can be generically a total number of $\binom{15}{3}+15\cdot 14 + 15 = \binom{17}{3} = 680$ terms in the intersection polynomial which can remain independent. However, 215 of those are nullified due to \cref{eq:intpoly-h11=15-1} and the remaining 465 terms are non-zero. However, our current aim is not to present the full intersection polynomial, but only to illustrate that the divisor topology analysis using NIDs is consistent with the database in \cite{Carta:2020ohw} where this example has been shown to indeed possess three possible orientifolds which can give $h^{1,1}_- = 7$.

\section{Moduli stabilization with rigidified divisors}
\label{sec:modustabrigdiv}

In standard type IIB flux compactifications with $F_3$ and $H_3$ fluxes, the tree-level superpotential depends only on the complex structure moduli $(U^i)$ and the axio-dilaton $(S)$, i.e. $W = W_0(U^\alpha, S)$. Subsequently, this flux-dependent superpotential can fix all complex structure moduli and the axio-dilaton supersymmetrically at leading order by enforcing:
\begin{equation}
D_\alpha W_0 = 0 =  D_{\ov{\alpha}} \, \ov{W}_0\coma \text{and}\quad D_S W_0 \, = 0 = D_{\ov{S}} \, \ov{W}_0\fstop
\label{UStab}
\end{equation}
The K\"ahler moduli can appear in $W$ only via non-perturbative effects. In what follows, we shall assume $n$ non-perturbative contributions to $W$ which can be generated by either rigid divisors, such as shrinkable dP 4-cycles, or non-rigid divisors with non-zero magnetic fluxes \cite{Bianchi:2011qh, Bianchi:2012pn, Louis:2012nb}. The corresponding non-perturbative superpotential is then:\footnote{The exponents $(- i\, a_i\, T_i)$ in \eqref{eq:Wnp-n} follow from the definition of the chiral variables, which have been chosen such that to make explicit the T-duality transformations between type IIA and type IIB \cite{Shukla:2019wfo}.}
\begin{equation}
W= W_0 + \sum_{i = 1}^n \, A_i\, e^{- i\, a_i\, T_i}\,.
\label{eq:Wnp-n}
\end{equation}
Note that in \eqref{eq:Wnp-n} there is no sum in the exponents $(- i\, a_i\, T_i)$, and summations are to be understood only when upper indices are contracted with lower indices; otherwise we will write an explicit sum as in \eqref{eq:Wnp-n}. We will suppose that out of $h^{1,1}$ K\"ahler moduli, only the first $n$ appear in $W$, i.e. $i=1,\ldots,n\leq h^{1,1}$.\footnote{We are assuming that $h^{1,1}_-=0$ in this section.}

For a generic superpotential which depends on all closed string chiral variables, namely $S$, $T_i$ and $U^\alpha$, the $F$-term scalar potential can be rewritten as:
\begin{equation}
\label{eq:Vgen0}
e^{- {\cal K}} \, V \, = (D_{\cal I} W) \, {\cal K}^{{\cal I}\ov{\cal J}} \, (\ov{D}_{\cal J} \ov{W}) - 3 |W|^2\coma
\end{equation}
where the summations over ${\cal I}$ and ${\cal J}$ run over all the moduli. However, assuming that the $S$ and $U$ moduli are stabilized as in \eqref{UStab} and considering a superpotential given by \eqref{eq:Wnp-n}, the scalar potential \eqref{eq:Vgen0} reduces to:
\begin{equation}
\begin{split}
V = & \, e^{{\cal K}} \, \left[ K_{T_i} K^{{T_i} \ov{T}_j}  W \ov{W}_{\ov{T}_j} + W_{T_i} K^{{T_i} \ov{T}_j} \left(\ov{W}_{\ov{T}_j}+ K_{\ov{T}_j} \ov{W} \right)+\right. \\
    &\left. + \left(K_{T_i} \, K^{{T_i} \ov{T}_j} \,  K_{\ov{T}_j} -3 \right)|W|^2 \right]\,.
\end{split}
\end{equation}
Moreover, this scalar potential can be written in a so-called master formula as the sum of the following three terms \cite{AbdusSalam:2020ywo}:
\begin{equation}
V = V_{\mathcal{O}(\alpha'^3)} + V_{\rm np1} + V_{\rm np2} \coma
\label{eq:Vgen-nGen}
\end{equation}
where
\begin{align}
V_{\mathcal{O}(\alpha'^3)} = & \, e^{{\cal K}} \, \frac{3 \, \hat\xi (\vo^2 + 7\,\vo\, \hat\xi +\hat\xi^2)}{({\cal V}-\hat{\xi }) (2\vo + \hat{\xi })^2}\, \,|W_0|^2\,, \nonumber \\
V_{\rm np1}  = & \,e^{{\cal K}}\, \sum_{i =1}^n \, 2 \, |W_0| \, |A_i|\, e^{- a_i \tau_i}\, \cos(a_i\, \rho_i + \theta_0 - \phi_i) \nonumber\\
 & \times ~\left[\frac{(4 \vo^2 + \vo \, \hat\xi+ 4\, \hat\xi^2)}{ (\vo - \hat\xi) (2\vo + \hat\xi)}\, (a_i\, \tau_i) +\frac{3 \, \hat\xi (\vo^2 + 7\,\vo\, \hat\xi + \hat\xi^2 )}{(\vo-\hat\xi) (2\vo + \hat\xi)^2}\right]\,,  \label{MasterF}\\
V_{\rm np2} =& \,e^{{\cal K}}\, \sum_{i=1}^n\,  \sum_{j=1}^n \, |A_i|\, |A_j| \, e^{-\, (a_i \tau_i + a_j \tau_j)} \, \cos(a_i\, \rho_i - a_j \, \rho_j -\phi_i + \phi_i)\times\nonumber  \\
 & \times \left[ -4 \left({\cal V}+\frac{\hat\xi}{2}\right) \, (\kappa_{ijk}\,t^k) \, a_i\, a_j\, + \frac{4{\cal V} - \hat{\xi}}{(\vo - \hat\xi)} \left(a_i\, \tau_i) \, (a_j\,\tau_j \right)  +\right.\nonumber\\
&  + \left.\,\frac{(4\vo^2  + \vo \, \hat{\xi} + 4\, \hat{\xi}^2)}{(\vo - \hat{\xi}) (2\vo + \hat{\xi})}\, (a_i\, \tau_i +a_j\, \tau_j) +\frac{3 \, \hat{\xi} (\vo^2 + 7\,\vo\, \hat{\xi}  +\hat\xi^2)}{({\cal V}-\hat{\xi }) (2\vo + \hat{\xi })^2}\right]\fstop\nonumber
\end{align}
Here we have introduced some model dependent parameters such as 
\begin{equation}
\tau_i = \partial_i{\cal V}\coma  {\cal V} = \frac{1}{6} \kappa_{ijk} t^i t^j t^k\coma 
W_0=|W_0|\, e^{i \, \theta_0}\coma  A_i = |A_i|\, e^{i\, \phi_i}\coma  \hat{\xi} = -\frac{\chi(CY) \zeta(3)}{16 \, \pi^3 \, g_s^{3/2}}\fstop
\end{equation} 
It was already demonstrated in \cite{AbdusSalam:2020ywo} how to use this master formula for generic moduli stabilization using CY from the KS database. One simply need to give some model dependent parameters and subsequently the problem boils down to minimize a function of multi-variables, which can be tackled with multiple ways, including the recent techniques of Machine Learning and Genetic Algorithm.

\subsection{A template using $h^{1,1}({\rm pCICY}) = 1$}

In this section, we present a procedure for performing  systematic moduli stabilization using the CY 3-folds with $h^{1,1} = h^{1,1}_+ = 1$ and assuming that the superpotential terms are generated by appropriate rigidification of the divisors. This procedure can be algorithmically adopted for analyzing the physical AdS/dS vacua in models of large number of K\"ahler moduli. So, let us begin with considering a KKLT like moduli stabilization using pCICY orientifolds.

\begin{table}[!htp]
\centering
\renewcommand{\arraystretch}{1.2}
\begin{tabular}{|Sc||Sc|Sc||Sc|Sc|Sc|} 
\hline
\shortstack{Model.\\\#} & \shortstack{pCICY\\$\{7820,7890\}$} & \makecell[c]{\vspace{0.3cm}$\{h^{1,1}, h^{2,1}\}$}  & \shortstack{Divisor\\Topologies} & \shortstack{Intersection\\polynomial}  & \makecell[c]{\vspace{0.1cm}$\displaystyle{\int_{_{\rm CY}} c_2 \wedge J_i}$} \\
\hhline{|=#=|=#=|=|=|}
$M_{1,1}$ & (7791, 7861) & $\{1, 65\}$ & T11 & $16 J_1^3$ & 64 \\
$M_{1,2}$ & (7808, 7878) & $\{1, 73\}$ & T9 & $9 J_1^3$ & 54 \\
$M_{1,3}$ & (7809, 7879) & $\{1, 73\}$ & T10 & $12 J_1^3$ & 60 \\
$M_{1,4}$ & (7819, 7889) & $\{1, 89\}$ & T8 & $8 J_1^3$ & 56\\
$M_{1,5}$ & (7820, 7890) & $\{1, 101\}$& T6 & $5 J_1^3$ & 50 \\
\hline
\end{tabular}
\caption{Topological data for favorable pCICYs with $h^{1,1} = 1$. Here, the Hodge number details for the divisor topologies Ti's are given in Table \ref{tab_divisor-topologies}.}
\label{tab_data-h11=1}
\end{table}

\noindent
In this context, the first example has been considered for moduli stabilization purpose in \cite{Bobkov:2010rf} while the last example, i.e. the Quintic CY 3-fold, has been considered in many occasions. Moreover, the Quintic CY has been also studied for $\alpha^\prime$ uplifted de Sitter vacua in \cite{Westphal:2006tn}. As a template for moduli stabilization, using the topological data for the five pCICY 3-folds given in Table \ref{tab_data-h11=1}, and considering the superpotential of the form,
\begin{equation}
W= W_0 + \, A_1\, e^{- i\, a_1\, T_1}\coma
\end{equation}
we will analyze the scalar potential, leading to AdS/dS vacua which we elaborate now.

\subsubsection{KKLT AdS vacua with anti-brane uplift}

In the absence of any $\alpha'$-corrections, the scalar potential of the simplest KKLT model with a single K\"ahler modulus can be given as,
\begin{equation}
V_{\rm KKLT} = \frac{9 \kappa_{111} g_s\, e^{K_{\rm cs}}}{\tau_1^2}\,a_1 |A_1|\, e^{-a_1 \tau_1} \left[\frac{|A_1|}{3} \, e^{-a_1 \tau_1} \left(a_1\, \tau_1+3 \right) - |W_0|\right]\coma
\label{VKKLT}
\end{equation}
where we have set the axion at its minimum defined by $(a_1 \rho_1 + \theta_0 - \phi_1) = \pi$ and used $\la s \ra = g_s^{-1}$. This leads to supersymmetric AdS vacua in the minimal KKLT model, which can be subsequently uplifted to non-supersymmetric dS vacua via various possible sources \cite{Kachru:2003aw,Burgess:2003ic}. Such an uplifting term can be schematically written as:
\begin{equation}
V_{\rm KKLT}^{\rm up} =  V_{\rm KKLT} + V_{\rm up}\qquad \text{with}\qquad V_{\rm up} = \frac{\delta}{t_1^p}\,,
\label{eq:kklt-up}
\end{equation}
where $\delta > 0$ is a tunable flux dependent parameter, while the parameter $p$ can be taken as $4 \leq p \leq 6$, which basically results in two main scenarios. The first one corresponds to $p = 6$ for placing an $\ov{D3}$-brane in the bulk, or a $D$-term from $D7$-brane with fluxes, and the second one corresponds to $p = 4$ for the $\ov{D3}$-brane located in a highly warped throat. As we will see from our numerical results in a moment, the need for quite tuned values of $\delta$ suggests that the warp throat scenario with $p = 4$ is more suitable for our case. Nevertheless, using the topological data from Table \ref{tab_data-h11=1} and the scalar potentials \eqref{VKKLT} and \eqref{eq:kklt-up}, we present the AdS as well as the uplifted dS vacua for all 5 pCICY 3-folds with $h^{1,1}=1$, with details as collected in Table \ref{tab_min-h11eq1a} and Table \ref{tab_min-h11eq1b} corresponding to the case $p = 4$ and $p = 6$ respectively.

\begin{table}[!htp]
\centering
\renewcommand{\arraystretch}{1.3}
\begin{tabular}{|c||c|c|c||c|c|c|c|} 
\hline
Model & $\la t_1\ra$ & $\la\vo\ra$ & $-V_0 \cdot 10^{14}$ & $\delta \cdot 10^{11}$ & $\la t_1\ra$& $\la\vo\ra$ &$V_0^{\rm up} \cdot 10^{18}$ \\
\hhline{|=#=|=|=#=|=|=|=|}
$M_{1,1}$ & 3.76811 & 142.672 & 6.35489 & 1.29 & 3.78147  & 144.195 & 4.51049  \\
$M_{1,2}$ & 5.02414 & 190.229 & 3.57463 & 2.30 & 5.04196 & 192.260 & 105.697  \\
$M_{1,3}$ & 4.35104 & 164.743 & 4.76617 & 1.72 & 4.36646 & 166.502 & 3.38287 \\
$M_{1,4}$ & 5.32891 & 201.769 & 3.17744 & 2.58 & 5.34781 & 203.923 & 2.25524 \\
$M_{1,5}$ & 6.74060 & 255.220 & 1.98590 & 4.13 & 6.76450 & 257.944 & 10.9614 \\
\hline
\end{tabular}
\caption{KKLT vacua using the scalar potentials potentials \eqref{VKKLT} and \eqref{eq:kklt-up} before and after the anti-brane uplifting corresponding to $p = 4$, and for all the pCICY 3-folds with $h^{1,1}= h^{1,1}_+ =1$. The underlying parameters are set as $W_0 = - 10^{-4}$, $a_1 = 0.1$, $g_s = 0.1$, $K_{\rm cs} = 0.1$, $A_1 = 1$ and the axion $\rho_1$ is minimized at $\langle\rho_1\rangle = 0$.}
\label{tab_min-h11eq1a}
\end{table}

\begin{table}[!htp]
\centering
\renewcommand{\arraystretch}{1.3}
\begin{tabular}{|c||c|c|c||c|c|c|c|} 
\hline
Model & $\la t_1\ra$ & $\la\vo\ra$ & $-V_0 \cdot 10^{14}$ & $\delta \cdot 10^{10}$ & $\la t_1\ra$& $\la\vo\ra$ &$V_0^{\rm up} \cdot 10^{17}$ \\
\hhline{|=#=|=|=#=|=|=|=|}
$M_{1,1}$ & 3.76811 & 142.672 & 6.35489 & 1.85 & 3.78946  & 145.111 & 5.66290  \\
$M_{1,2}$ & 5.02414 & 190.229 & 3.57463 & 5.85 & 5.05262 & 193.482 & 5.04044  \\
$M_{1,3}$ & 4.35104 & 164.743 & 4.76617 & 3.29 & 4.37569 & 167.560 & 5.83016 \\
$M_{1,4}$ & 5.32891 & 201.769 & 3.17744 & 7.40 & 5.35911 & 205.219 & 2.83145 \\
$M_{1,5}$ & 6.74060 & 255.220 & 1.98590 & 18.93 & 6.77880 & 259.583 & 0.326836 \\
\hline
\end{tabular}
\caption{KKLT vacua using the scalar potentials potentials \eqref{VKKLT} and \eqref{eq:kklt-up} before and after the anti-brane uplifting corresponding to $p = 6$, and for all the pCICY 3-folds with $h^{1,1}= h^{1,1}_+ =1$. The underlying parameters are set as $W_0 = - 10^{-4}$, $a_1 = 0.1$, $g_s = 0.1$, $K_{\rm cs} = 0.1$, $A_1 = 1$ and the axion $\rho_1$ is minimized at $\langle\rho_1\rangle = 0$.}
\label{tab_min-h11eq1b}
\end{table}

\subsubsection{Stable/tachyonic dS vacua using $\alpha^\prime$-uplift}

A less simple task will be to use the generic BBHL corrected single field scalar potential \cite{Becker:2002nn}, i.e. without setting the $\alpha'$ corrections to zero. In this case, after fixing the axionic moduli at its minimum via $(a_1 \rho_1 + \theta_0 - \phi_1) = \pi$, the saxion dependent scalar potential can be given as a sum of the following 3 terms:
\begin{align}
V_{\mathcal{O}(\alpha'^3)} = &\,  e^{{\cal K}} \, \frac{3 \, \hat\xi (\vo^2 + 7\, \hat\xi \,\vo +\hat{\xi}^2)}{(\vo-\hat\xi) (2\vo+ \hat\xi)^2}\, \,|W_0|^2\coma \nonumber\\
V_{\rm np1} =& - 2\, e^{{\cal K}}\,|W_0| \, |A_1|\, e^{- a_1 \tau_1} \left[\frac{(4\vo^2  + \vo \, \hat\xi + 4\, \hat{\xi}^2)}{(\vo - \hat\xi) (2 \vo + \hat\xi)}\, (a_1 \tau_1) +\frac{3 \, \hat\xi (\vo^2+ 7\, \hat\xi \,\vo +\hat{\xi}^2 )}{(\vo-\hat\xi) (2\vo + \hat\xi )^2}\right ]\coma \\
V_{\rm np2} =&\, e^{{\cal K}}\,  \, |A_1|^2 \, e^{- 2 a_1 \tau_1} \left[ -4\, a_1^2 \left(\vo+\frac{\hat{\xi}}{2}\right) \, \sqrt{2\, \kappa_{111}\,\tau_1}  + \frac{4\vo - \hat\xi}{(\vo - \hat\xi)} (a_1 \tau_1)^2+\right. \nonumber \\
& + \left.\frac{(4 \vo^2 + \vo \, \hat{\xi} + 4\, \hat{\xi}^2)}{ (\vo - \hat\xi) (2 \vo + \hat\xi)}\, (2\,a_1\,\tau_1) +\frac{3 \, \hat\xi (\vo^2  + 7\, \hat\xi \,\vo + \hat{\xi}^2)}{(\vo-\hat\xi) (2 \vo + \hat\xi)^2}\right]\fstop \nonumber
\end{align}
This scalar potential can be minimized numerically with respect to either $t_1$ or $\tau_1$, since in this simple case the conversion between 2- and 4-cycle moduli is trivial.  As we will show later on, we find some stable dS vacua which turn out to be realized at quite lower volume level, and one may find its stability to be weak against other sub-leading corrections.

\subsubsection*{Tachyonic dS vacua with larger volume}
We find that there can be tachyonic dS vacua for relatively larger volume ${\cal V} \simeq {\cal O}(100)$ while using $a_i$'s still not too low. These are given in \cref{tab_alphamin-h11eq1,tab_alphamin-h11eq1-05}.

\begin{table}[!htp]
\centering
\renewcommand{\arraystretch}{1.3}
\begin{tabular}{|c||c|c|c|c|c|c|} 
\hline
Model & $\la t_1\ra$ & $\la\vo\ra$ & $\la\hat\xi\ra$ & $V_0 \cdot 10^8$ & Eigenvalues($\la V_{ij} \ra$)\\
\hhline{|=#=|=|=|=|=|=|}
$M_{1,1}$  & 3.31014 & 96.7177  & 1.49783 & 0.14949 & $\{-2.33705\cdot 10^{-6}, \, 1.12791 \cdot 10^{-9}\}$ \\
$M_{1,2}$  & 4.48991 & 135.770  & 1.68506 & 6.12546 & $\{-5.37276\cdot 10^{-7},\, 4.36364\cdot 10^{-10}\}$\\
$M_{1,3}$  & 3.83263 & 112.595  & 1.68506 & 0.10671 & $\{-1.25072\cdot 10^{-6},\, 7.97739\cdot 10^{-10}\}$ \\
$M_{1,4}$  & 4.69500 & 137.989  & 2.05952  & 7.08644 & $\{-5.53698\cdot 10^{-7},\, 5.29375\cdot 10^{-10}\}$ \\
$M_{1,5}$  & 6.00346 & 180.312 & 2.34036 & 3.62639 & $\{-1.76802\cdot 10^{-7},\, 2.61241\cdot 10^{-10}\}$ \\
\hline
\end{tabular}
\caption{For all CICYs with $h^{1,1}=1$, we present the $\alpha^\prime$-uplifted tachyonic dS vacua lying in a relatively larger volume regime. The underlying parameters are set as $W_0 = -1, g_s = 0.35, a_1 = \frac{\pi}{16}$, $K_{\rm cs} = 1$, $A_1 = 1$ and the axion $\rho_1$ is minimized at $\la\rho_1\ra = 0$.}
\label{tab_alphamin-h11eq1}
\end{table}

\begin{table}[!htp]
\centering
\renewcommand{\arraystretch}{1.3}
\begin{tabular}{|c|c||c|c|c|c|c|c|} 
\hline
\# & $g_s$ & $\la t_1\ra$ & $\la\vo\ra$ & $\la\hat\xi\ra$ & $V_0 \cdot 10^8$ & Eigenvalues($\la V_{ij} \ra$)\\
\hhline{|=|=#=|=|=|=|=|}
1 & 0.20 & 5.39806 & 131.079  & 5.41802 & 0.11977  & $\{-5.89327\cdot 10^{-7}, \, 1.27980 \cdot 10^{-9}\}$ \\
2 & 0.25 & 5.67204 & 152.068  & 3.87682 & 6.99281 & $\{-3.43508\cdot 10^{-7}, \, 6.15255 \cdot 10^{-10}\}$ \\
3 & 0.30 & 5.86049 & 167.734  & 2.9492 & 4.81623 & $\{-2.35651\cdot 10^{-7}, \, 3.76577 \cdot 10^{-10}\}$ \\
4 & 0.35 & 6.00346 & 180.312 & 2.34036 & 3.62639 & $\{-1.76802\cdot 10^{-7}, \, 2.61241 \cdot 10^{-10}\}$ \\
\hline
\end{tabular}
\caption{String coupling variation for the $\alpha^\prime$-uplifted tachyonic dS vacua of the Quintic 3-fold $M_{1,5}$. The underlying parameters are set as $W_0 = -1, a_1 = \frac{\pi}{16}$, $K_{\rm cs} = 1$, $A_1 = 1$ and the axion $\rho_1$ is minimized at $\la\rho_1\ra = 0$.}
\label{tab_alphamin-h11eq1-05}
\end{table}

\subsubsection*{Stable dS vacua with smaller volume}

The results of our numerical analysis are presented in \cref{tab_alphamin-h11eq1-1,tab_alphamin-h11eq1-2}. The former of these, i.e. Table \ref{tab_alphamin-h11eq1-1}, shows the presence of dS vacua located at relatively lower volume, and therefore the effective field theory can only be marginally under control given that volume ${\cal V}$ is not large enough. A relatively large rank of the gauge group ($N$) appearing in the exponent of the non-perturbative effect is needed to obtain a minimum where the volume can get quite large, which is difficult but maybe feasible, e.g. see \cite{Louis:2012nb}, where an explicit example with $N=24$ in a globally consistent CY orientifold model was constructed. In this regard, a set of larger values of $\la \vo\ra \simeq \mathcal{O}(100)$ realized using $N\sim\mathcal{O}(100)$ and $|W_0| \sim \mathcal{O}(1-25)$ \cite{Balasubramanian:2004uy, Westphal:2006tn, Rummel:2011cd, Louis:2012nb} are illustrated in Table \ref{tab_alphamin-h11eq1-2}, even though these values have their own limitations and criticism \cite{Cicoli:2013swa, Conlon:2012tz}.

\begin{table}[!htp]
\centering
\renewcommand{\arraystretch}{1.3}
\begin{tabular}{|c||c|c|c|c|c|c|c|} 
\hline
Model & $g_s$ & $-W_0$ & $\la t_1\ra$ & $\la\tau_1\ra$ & $\la\vo\ra$ & $\hat\xi$ & $V_0 \cdot 10^6$ \\
\hhline{|=#=|=|=|=|=|=|=|}
$M_{1,1}$ & 0.2 & 2.30 & 1.97526 & 31.2133  & 20.5515 & 3.46753 & 7.12393 \\
$M_{1,2}$ & 0.2 & 2.90 & 2.61562 & 30.7865 & 26.8419 & 3.90097 & 2.55535 \\
$M_{1,3}$ & 0.2 & 2.37 & 2.26773 & 30.8557 & 23.3242 & 3.90097 & 1.47466 \\
$M_{1,4}$ & 0.2 & 2.38 & 2.78034 & 30.9213 & 28.6572 & 4.76786 & 1.56166 \\
$M_{1,5}$ & 0.2 & 2.76 & 3.50740 & 30.7546  & 35.9563 & 5.41802 & 0.847185 \\
\hline
\end{tabular}
\caption{For all CICYs with $h^{1,1}=1$, we present the $\alpha'$-uplifted stable dS vacua lying in a relatively lower volume regime. The underlying parameters are set as $a_1 = 0.1$, $K_{\rm cs} = 1$, $A_1 = 1$ and the axion $\rho_1$ is minimized at $\la\rho_1\ra = 0$.}
\label{tab_alphamin-h11eq1-1}
\end{table}

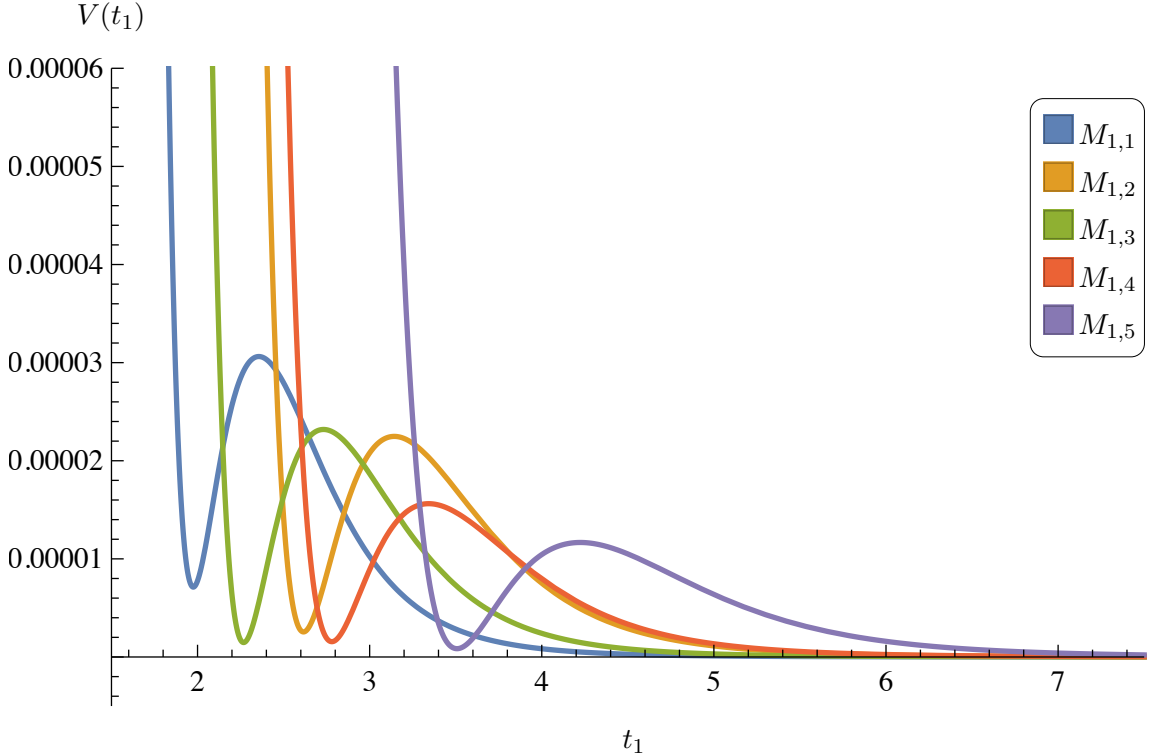
\begin{figure}[!htp]
\begin{center}
\begin{tikzpicture}[scale=1]
\node (plt) at (0,0) {\includegraphics[width=\textwidth,keepaspectratio]{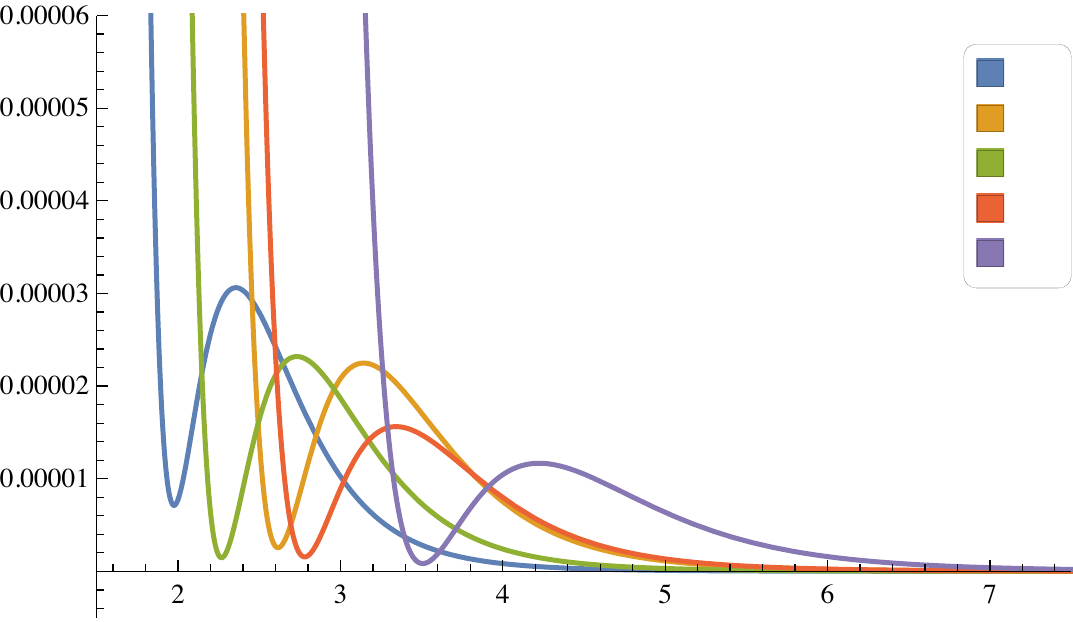}};
\node[above= of plt,node distance=0cm,yshift=-1cm,xshift=-6.2cm] {$V(t_1)$};
\node[below= of plt,node distance=0cm,yshift=1cm,xshift=0.75cm] {$t_1$};
\node[right= of plt,node distance=0cm,yshift=3.23cm,xshift=-2.17cm] {$M_{1,1}$};
\node[right= of plt,node distance=0cm,yshift=2.59cm,xshift=-2.17cm] {$M_{1,2}$};
\node[right= of plt,node distance=0cm,yshift=1.96cm,xshift=-2.17cm] {$M_{1,3}$};
\node[right= of plt,node distance=0cm,yshift=1.33cm,xshift=-2.17cm] {$M_{1,4}$};
\node[right= of plt,node distance=0cm,yshift=0.7cm,xshift=-2.17cm] {$M_{1,5}$};
\end{tikzpicture}
\end{center}
\caption{KKLT scalar potentials $V(t_1)$ with $\alpha'$-uplift for all 5 models in Table \ref{tab_data-h11=1}. The corresponding details about the moduli VEVs and $V_0$ at the minimum are given in Table \ref{tab_alphamin-h11eq1-1}.}
\label{fig_upKKLT}
\end{figure}

\begin{table}[!htp]
\centering
\renewcommand{\arraystretch}{1.3}
\begin{tabular}{|c||c|c|c|c|c|c|c|} 
\hline
Model & $g_s$ & $-W_0$ & $\la t_1\ra$ & $\la\tau_1\ra$ & $\la\vo\ra$ & $\hat\xi$ & $V_0 \cdot 10^8$ \\
\hhline{|=#=|=|=|=|=|=|=|}
$M_{1,1}$ & 0.1 & 5.89 & 3.43588 & 94.4423  & 108.164 & 9.80766 & 1.38237 \\
$M_{1,2}$ & 0.1 & 7.35 & 4.55355 & 93.3067  & 141.626 & 11.0336 & 0.74186 \\
$M_{1,3}$ & 0.1 & 6.10 & 3.96620 & 94.3846  & 124.783 & 11.0336 & 3.58413 \\
$M_{1,4}$ & 0.2 & 22.15 & 4.68009 & 87.6131  & 136.679 & 4.76786 & 3.83906 \\
$M_{1,5}$ & 0.2  & 25.10 & 5.90179 & 87.0779 & 171.305 & 5.41802 & 8.39116 \\
\hline
\end{tabular}
\caption{For all CICYs with $h^{1,1}=1$, we present the $\alpha'$-uplifted stable dS vacua lying in a relatively larger volume regime. The underlying parameters are set as $a_1 = \frac{\pi}{100}$, $K_{\rm cs} = 1$, $A_1 = 1$ and the axion $\rho_1$ is minimized at $\la\rho_1\ra = 0$.}
\label{tab_alphamin-h11eq1-2}
\end{table}


\subsection{A lower bound on the CY volume ${\cal V}$}

In our detailed analysis of the divisor topologies and the other ingredients such as CY volume ${\cal V}$ as well as the 2- and 4-cycle volume moduli, we make some interesting observations. In fact, given that the triple intersection numbers $\kappa_{ijk}$ always turn out to be non-negative for all the favorable pCICYs in the database, we observe that the volume of the CY has a huge number of terms (especially for larger $h^{1,1}$) which are guaranteed to be positive if the divisor basis corresponds to the K\"ahler cone generators\footnote{In a recent work of perturbatively flat vacua analysis \cite{Carta:2021kpk}, the utility of K\"ahler cone basis over the other possible divisor basis has been argued.} which would simply mean $t_i > 0$ for all $i \in h^{1,1}_+$(CY). Moreover, for the validity of the supergravity approximations in the sense of the EFT description, we would need moduli VEVs such that $t_i > 1 \, \, \forall i$. Subsequently, this naive assumption, which is anyway the core of any model building phenomenology, results in some interesting lower bound for the volume ${\cal V}$ of the CY, and this is irrespective of the choice of moduli stabilization scheme. For having some estimates we considered all the 4874 CICYs in the favorable list in \cite{Anderson:2017aux} for which K\"ahler cone can be simply given as $t_i > 0$. Subsequently, imposing that $t_i = 1$ we get some rough estimates for ${\cal V}_{\rm min}$ for each of such CYs, and we observe that maximum value which correspond to polytope ID 232 of 7890 collection (and 193 in the favorable CICY collection) can result ${\cal V}$ as large as 684 by simply assuming $t_i =1$. So without doing any moduli stabilization, one confirms that if the dynamical minimization has to lie withing a trustworthy EFT regime, the volume for this model cannot be less than 684. In fact this value can go as high as 85500 if one wants, e.g., $t_i = 5$. This observation is interesting and encouraging to explore such CICYs with larger $h^{1,1}$ values. In fact, we observed that the very first example in the list which has 465 terms in ${\cal V}$ can lead to ${\cal V}_{\rm min} = 1404$ by merely imposing $t_i = 1$, however given that the explicit K\"ahler cone conditions for this CY is not guaranteed to be $t_i > 0$ we exclude this example while estimating the max/min values of ${\cal V}$. Some estimates are given in Table \ref{tab:CICYsvolume} in this regard.

\begin{table}[!htp]
    \centering
    \renewcommand{\arraystretch}{1.3}
    \begin{tabular}{|c|c|c|c||c|c|}
    \hline
        \# CICY & \# fav. CICY &  $\left\{h^{1,1},h^{2,1}\right\}$ & $t_i$ & ${\cal V}$\\
        \hhline{|=|=|=|=#=|=|}
        \multirow{5}{*}{$232$} & \multirow{5}{*}{$193$} & \multirow{5}{*}{$\{12,18\}$}   & $1$ & $684$ \\
        \cline{4-6}
                                &                       &                                & $2$ & $5472$ \\
        \cline{4-6}
                                &                       &                                & $3$ & $18468$ \\
        \cline{4-6}
                                &                       &                                & $4$ & $43776$ \\
        \cline{4-6}
                                &                       &                                & $5$ & $85500$ \\
    \hhline{|=|=|=|=#=|=|}
        \multirow{5}{*}{$7890$} & \multirow{5}{*}{$7820$} & \multirow{5}{*}{$\{1,101\}$} & $1$ & $\frac{5}{6}$ \\
        \cline{4-6}
                                &                           &                              & $2$ & $\frac{20}{3}$ \\
        \cline{4-6}
                                &                           &                              & $3$ & $\frac{45}{2}$ \\
        \cline{4-6}
                                &                           &                              & $4$ & $\frac{160}{3}$ \\
        \cline{4-6}
                                &                           &                              & $5$ & $\frac{625}{6}$ \\
    \hline
    \end{tabular}
    \caption{Estimates for the maximum and minimum values of the CY volumes (${\cal V}$) among the K\"ahler positive CICYs for different values of $t_i$.}
    \label{tab:CICYsvolume}
\end{table}


\section{Conclusions}
\label{sec:concl}

The available dataset of the projective complete intersection Calabi-Yau 3-folds (pCICYs) has a huge number of interesting examples which have remained unexplored for phenomenological applications. With the recent attraction along these lines in \cite{Carta:2020ohw,Carta:2021sms}, we take some further steps to promote this pCICYs database for moduli stabilization and other subsequent phenomenological purposes.  

In this regard, first we have computed all the divisor topologies as the same play significant role in attempts for constructing concrete global models. To our surprise we have found that there are only 11 distinct topologies for the so-called coordinate divisors of the favorable pCICY 3-fold list which consists of a total of 7820 spaces having 57885 possible divisors with distinct GLSM charges. We also observe that it would be easier to have explicit CICY orientifold construction using exchange involution, which could result in non-trivial $(1,1)$-cohomology of the respective 3-folds. It is because of the frequent presence of the so-called ``non-trivially identical divisors" (NIDs) which are exchanged under the involution. We find that only 111 pCICY 3-folds (out of 7820) do not have such NIDs and hence one would usually not expect to have $h^{1,1}_-({\rm pCICY})$ = 0 for such spaces.

Having all the divisor topologies for the pCICY database, we wanted to look for the so-called ample divisors as the same have been proposed to stabilize all the $h^{1,1}_+$ saxions of the K\"ahler moduli by a single superpotential term \cite{Bobkov:2010rf}. In this regard, we listed the necessary conditions for ample divisors and subsequently explored the dataset of the favorable examples. We find an interesting condition: $6\chi_{_h} < \chi < 12 \chi_{_h}$ which an ample divisor need to satisfy along with others listed in Eq. \eqref{eq:ample-1}. This condition is quite restrictive if one wants an ample divisor to be rigid as well, and leaves only five possibilities using $\chi_{_h} = 1$. In addition, we find that ample divisors as we propose in Eq. \eqref{eq:ample-sum} are never rigid and hence in order to start moduli stabilization using such divisors one would need the rigidification of ample divisors e.g., a la techniques developed in \cite{Bianchi:2011qh,Bianchi:2012pn, Louis:2012nb} which can be an interesting aspect to explore. 

To illustrate the strategy about the possibility of directly adopting the pCICY dataset for phenomenological purposes, we have performed a systematic approach for the moduli stabilization using the concrete topological data of the pCICY 3-folds for $h^{1,1} = 1$. We have demonstrated the usual KKLT-like AdS and its uplifting through anti-D3 brane for all the five examples with $h^{1,1} =1$. Moreover, using the $\alpha^\prime$-corrections we have illustrated the uplifted dS vacua which come in two categories; one which corresponds to a minimum is realized at $\la {\cal V} \ra \sim {\cal O}(30)$ while the other one which corresponds to a tachyonic dS solution is realized at $\la {\cal V} \ra \sim {\cal O}(200)$. In fact, we also show that it is possible to realize $\la {\cal V} \ra \sim {\cal O}(100-200)$ using non-perturbative superpotentials arising from gaugino condensation with large rank of the gauge group.

We expect that having all the topological ingredients at hand, it should be possible to perform the moduli stabilization using many K\"ahler moduli in explicit orientifolds of pCICYs. Given the sizable number of examples, the dataset can be useful for performing a systematic search of AdS/dS vacua using contemporary techniques such as Machine Learning, Neural Network and Genetic Algorithms. To demonstrate these arguments let us consider the very first CY in the list of pCICYs which has $(h^{1,1}, h^{2,1}) = (15, 15)$, and hence $\chi = 0$ leading to $\hat\xi = 0$, i.e. the $(\alpha^\prime)^3$-corrections to the scalar potential are trivial. Subsequently, using a non-perturbative superpotential having 15 exponential terms of the form given in Eq. \eqref{eq:Wnp-n}, the master formula \cref{MasterF} for the scalar potential takes the following simple form,
\begin{equation}
\begin{split}
V  = & \,e^{{\cal K}}\, \sum_{i =1}^{15} \, 4 \, |W_0| \, |A_i|\, e^{- a_i \tau_i}\, (a_i\, \tau_i)\,  \cos(a_i\, \rho_i + \theta_0 - \phi_i)+  \\
& +  \,e^{{\cal K}}\, \sum_{i=1}^{15}\,  \sum_{j=1}^{15} \, 2\, |A_i|\, |A_j| \, e^{-\, (a_i \tau_i + a_j \tau_j)} \, \cos(a_i\, \rho_i - a_j \, \rho_j -\phi_i + \phi_j)\times  \\
 & \times \left[ -2 \, {\cal V} \, (\kappa_{ijk}\,t^k) \, a_i\, a_j\, + 2 \left(a_i\, \tau_i) \, (a_j\,\tau_j \right) + \, (a_i\, \tau_i +a_j\, \tau_j)\right].
\end{split}
\end{equation}
This form of scalar potential is easy to construct for numerical analysis by simply using triple intersection numbers $\kappa_{ijk}$, and subsequently the task remains to minimize a complicated multi-variable function. In this regard, we have also argued that if one is working with the basis of K\"ahler cone generators, then the physicality conditions on the two-cycle volumes can simply be imposed as $t_i > 0$ or even $t_i > 1$ for all $i\in h^{1,1}_+(CY)$, in order to trust the supergravity approximation in the sense of EFT description. In fact, while making this analysis we have realized that there can be a limit on the minimum value of the CY volume ${\cal V}$ which can be estimated by assuming $t_i = 1$. Such a bound is there due to the fact that all the triple intersection numbers we have found in this pCICY dataset are non-negative, and if the K\"ahler cone condition is such that $t_i > 0$: therefore, the estimated minimum value of the volume is irrespective of the moduli stabilization scheme, and only assumes the validity with EFT description. For small $h^{1,1}$ models, this does not appear to be relevant, and therefore it is not usually emphasized. For example, the simplest model of the Quintic CY has the volume given as ${\cal V} = \frac56 t_1^3$ and setting $t_1 =1$ through EFT requirement does not appear to give any significant bound. However, for CYs in which the volume form has 465 term (like the example with $h^{1,1} =15$ we discussed) and if each term is positive, then it is possible to realize quite large values of ${\cal V}$ whenever EFT validity is imposed, irrespective of which moduli stabilization mechanism is used. This information is also encouraging to explore the CYs with large number of K\"ahler moduli, and we hope to report back on some of these issues in a near future work.

\acknowledgments

The authors thank I\~naki Etxebarria, Xin Gao and Nicole Righi for useful discussions. F.C. is supported by STFC consolidated grant ST/T000708/1. The work of A.M. is supported in part by Deutsche Forschungsgemeinschaft under Germany's Excellence Strategy EXC 2121 Quantum Universe 390833306. P.S. is thankful to Paolo Creminelli, Atish Dabholkar and Fernando Quevedo for their support.

\appendix


\section{Divisor topologies of non-favorable pCICYs}
\label{app:non-favorable}

There are 70 CICY 3-folds out of the full collection of 7890 spaces which are non-favorable. These 70 spaces can be further classified into two categories, namely being product-type and non-product-type. It turns out that there are 22 spaces which are of product-type (PT) while the remaining 48 are of non-product-type (NPT).

\begin{table}[!htp]
\centering
\renewcommand{\arraystretch}{1.2}
\begin{tabular}{|Sc||Sc|Sc|} 
\hline
\makecell[c]{\vspace{0.3cm} $h^{1,1}$} & \shortstack{\# of non-fav\\ NPT pCICYs} & \shortstack{\# of divisors \\of NPT pCICYs} \\
 \hhline{|=#=|=|}
 1-6 & 0 & 0  \\
 7  & 1 & 2  \\   
 8  & 3 & 8  \\
 9  & 4 & 13  \\
 10  & 5 & 19  \\
 11  & 4 & 15  \\
 12  & 6 & 22  \\
 13  & 4 & 14  \\
 14  & 4 & 15 \\
 15  & 1 & 3 \\ 
 16  & 1 & 4 \\
 17-18  & 0 & 0 \\
 19 & 15 & 57 \\
  \hline
 Total \# & 48 & 172 \\
 \hline
\end{tabular}
\caption{Number of non-favorable NPT pCICYs and their divisors for $1\leq h^{1,1} \leq 19$.}
\label{tab_non-favorable-cicys-and-divisors}
\end{table}

\noindent
For the non-favorable NPT pCICYs, the number of independent divisors in the K\"ahler class is always less than $h^{1,1}$ of the CY 3-fold. We also observe that there are some divisors of the non-favorable pCICYs which have $h^{1,0}(D) \neq 0$. The details on the divisor topologies are summarized in Table \ref{tab_non-favorable-divisor-topologies}.

\begin{table}[!htp]
\centering
\renewcommand{\arraystretch}{1.2}
\begin{tabular}{|Sc||Sc|Sc|Sc|Sc|Sc|Sc|} 
\hline
\shortstack{Sr.\\\#} & \shortstack{Divisor \\ topology} & \shortstack{frequency\\(172 divisors)}  & \shortstack{frequency\\(48 spaces)} & \shortstack{$h^{1,1}$\\(pCICY)} & \makecell{\vspace{0.2cm} $\displaystyle{\int_{_{\rm CY}} \hat{D}^3}$} & \shortstack{NIDs\\(48 spaces)} \\
\hhline{|=#=|=|=|=|=|=|} 
T1 & $\{1, 0, 1, 20\}$ & 73 & 35 & 8-16, 19 & 0  & 24 \\
T2 & $\{1, 0, 2, 30\}$ & 27 & 23 & 7-14, 19 & 0 & 4 \\
T12 & $\{1, 1, 2, 22\}$ & 19 & 17 & 9-16, 19 & 0 & 2 \\
T13 & $\{1, 1, 3, 32\}$ & 19 & 17 & 8-15, 19 & 0 & 2 \\
T14 & $\{1, 1, 4, 42\}$ & 19 & 17 & 7-14, 19 & 0 & 2 \\
T15 & ${\mathbb T}^4 \equiv \{1, 2, 1, 4\}$ & 15 & 15 & 19 & 0 & 0 \\
\hline
\end{tabular}
\caption{Divisor topologies for non-favorable pCICYs and their frequencies of appearance. Here the Hodge numbers are collected as $\{h^{0,0}, h^{1,0}, h^{2,0}, h^{11}\}$.}
\label{tab_non-favorable-divisor-topologies}
\end{table}

\noindent
We make the following observation from the divisor topologies collected in Table \ref{tab_non-favorable-divisor-topologies}.
\begin{itemize}
\item There are in total six types of divisor topologies which appear in the non-favorable non-product-type pCICY spaces. It turns out that two of the most frequent topologies, namely T1 and T2, have been already part of the favorable pCICYs case, while the remaining four topologies are new as well as peculiar in the sense that $h^{1,0}(D) \in \{1, 2\}$ unlike the favorable pCICY 3-folds.

\item
We note that all the divisors of the non-favorable pCICYs (which are not of the product-type) have vanishing self triple intersection number as can been from the Hodge number condition \cref{eq:Dcube=0}.

\item
All the pCICYs with $h^{1,1} =19$ are known to be related by ineffective splittings. This means, they are just different ambient space descriptions of the same Calabi-Yau manifold, known as the Sch\"on \cite{Anderson:2017aux}. The divisors of the Sch\"on which are visible as pullbacks of ambient space divisors differ, in general, in the 15 realizations. Interestingly, we observe the presence of ${\mathbb T}^4$ surfaces as divisors. To our knowledge, this is the only observation of a CY 3-fold having the ${\mathbb T}^4$ surface as a coordinate divisor. Moreover, such an example was studied for moduli stabilization purposes in \cite{Bobkov:2010rf}. The presence of a ${\mathbb T}^4$ divisor can be understood also from the realization of the Sch\"on as the fibered product of two general elliptic surfaces, identified over a common $\mathbb{P}^1$.

\item
We notice that there is a non-trivial frequency of these divisor topologies to appear as NID pairs (except for the ${\mathbb T}^4$ topology case) and subsequently there is possibility of constructing orientifolds using exchange involutions so that to resut in non-trival $(1,1)$-cohomology of the respective pCICYs.

\end{itemize}


\section{Topological data for favorable pCICYs with $h^{1,1} = 2$}
\label{sec:topdatah112}

Our findings are promising for a flat vacua analysis similar to what has been recently done in \cite{Carta:2021kpk}, but this time, using pCICYs. The relevant data for $h^{1,1}=2$ can be found in Table \ref{tab_data-h11=2}, and we will classify all possible flat vacua that can be constructed using the techniques introduced in \cite{Demirtas:2019sip} in a coming up work \cite{Carta:2201aaaaa}. 

\begin{center}
\renewcommand{\arraystretch}{1.2}
  \begin{longtable}{|c||c|c||c|c|c|} 
 \caption{Topological data for favorable pCICYs with $h^{1,1} = 2$. The intersection numbers $\kappa_{ijk}$ are collected as $\{\kappa_{111}, \kappa_{112}, \kappa_{122}, \kappa_{222}\}$, while other details for topologies Ti's are given in Table \ref{tab_divisor-topologies}.}  \\
\hline
\shortstack{$M_{i,j}$\\$(i=h^{1,1})$} & \shortstack{Space \#\\$\{7820, 7890\}$}  & \makecell[c]{\vspace{0.3cm}$\{h^{1,1}, h^{2,1}\}$} & \shortstack{Topology\\of $J_i$} & \makecell[c]{\vspace{0.3cm}$\kappa_{ijk}$}  & \makecell[c]{\vspace{0.1cm}$\displaystyle{\int_{_{\rm CY}} c_2 \wedge J_i}$}\\
\hhline{|=#=|=#=|=|=|}
\endhead
\label{tab_data-h11=2}$M_{2,1}$ & $\{7573, 7643\}$ & $\{2, 46\}$ & \{T2, T9\} & $\{0, 4, 12, 8\}$ & $\{36, 56\}$ \\
$M_{2,2}$ & $\{7574, 7644\}$ & $\{2, 46\}$ & \{T7, T7\} & $\{4, 12, 12, 4\}$ & $\{52, 52\}$ \\
$M_{2,3}$ & $\{7598, 7668\}$ & $\{2, 47\}$ & \{T2, T5\} & $\{0, 3, 9, 6\}$ & $\{36, 48\}$ \\
$M_{2,4}$ & $\{7655, 7725\}$ & $\{2, 50\}$ & \{T2, T10\} & $\{0, 4, 12, 12\}$ & $\{36, 60\}$ \\
$M_{2,5}$ & $\{7656, 7726\}$ & $\{2, 50\}$ & \{T4, T9\} & $\{2, 8, 12, 8\}$ & $\{44, 56\}$ \\
$M_{2,6}$ & $\{7688, 7758\}$ & $\{2, 52\}$ & \{T2, T7\} & $\{0, 4, 10, 4\}$ & $\{36, 52\}$ \\
$M_{2,7}$ & $\{7689, 7759\}$ & $\{2, 52\}$ & \{T4, T7\} & $\{2, 8, 10, 4\}$ & $\{44, 52\}$ \\
$M_{2,8}$ & $\{7691, 7761\}$ & $\{2, 52\}$ & \{T6, T6\} & $\{5, 10, 10, 5\}$ & $\{50, 50\}$ \\
$M_{2,9}$ & $\{7729, 7799\}$ & $\{2, 55\}$ & \{T4, T4\} & $\{2, 7, 7, 2\}$ & $\{44, 44\}$ \\
$M_{2,10}$ & $\{7736, 7806\}$ & $\{2, 56\}$ & \{T1, T5\} & $\{0, 0, 6, 6\}$ & $\{24, 48\}$ \\
$M_{2,11}$ & $\{7737, 7807\}$ & $\{2, 56\}$ & \{T2, T9\} & $\{0, 4, 10, 8\}$ & $\{36, 56\}$ \\
$M_{2,12}$ & $\{7738, 7808\}$ & $\{2, 56\}$ & \{T2, T8\} & $\{0, 3, 9, 9\}$ & $\{36, 54\}$ \\
$M_{2,13}$ & $\{7739, 7809\}$ & $\{2, 56\}$ & \{T4, T6\} & $\{2, 7, 9, 5\}$ & $\{44, 50\}$ \\
$M_{2,14}$ & $\{7746, 7816\}$ & $\{2, 58\}$ & \{T1, T9\} & $\{0, 0, 8, 8\}$ & $\{24, 56\}$ \\
$M_{2,15}$ & $\{7747, 7817\}$ & $\{2, 58\}$ & \{T1, T10\} & $\{0, 0, 8, 12\}$ & $\{24, 60\}$ \\
$M_{2,16}$ & $\{7749, 7819\}$ & $\{2, 58\}$ & \{T1, T11\} & $\{0, 0, 8, 16\}$ & $\{24, 64\}$ \\
$M_{2,17}$ & $\{7751, 7821\}$ & $\{2, 58\}$ & \{T2, T6\} & $\{0, 4, 8, 5\}$ & $\{36, 50\}$ \\
$M_{2,18}$ & $\{7752, 7822\}$ & $\{2, 58\}$ & \{T1, T9\} & $\{0, 0, 8, 8\}$ & $\{24, 56\}$ \\
$M_{2,19}$ & $\{7753, 7823\}$ & $\{2, 58\}$ & \{T1, T11\} & $\{0, 0, 8, 16\}$ & $\{24, 64\}$ \\
$M_{2,20}$ & $\{7763, 7833\}$ & $\{2, 59\}$ & \{T2, T4\} & $\{0, 3, 7, 2\}$ & $\{36, 44\}$ \\
$M_{2,21}$ & $\{7770, 7840\}$ & $\{2, 62\}$ & \{T1, T8\} & $\{0, 0, 6, 9\}$ & $\{24, 54\}$ \\
$M_{2,22}$ & $\{7774, 7844\}$ & $\{2, 62\}$ & \{T2, T4\} & $\{0, 4, 6, 2\}$ & $\{36, 44\}$ \\
$M_{2,23}$ & $\{7783, 7853\}$ & $\{2, 64\}$ & \{T2, T7\} & $\{0, 4, 8, 4\}$ & $\{36, 52\}$ \\
$M_{2,24}$ & $\{7788, 7858\}$ & $\{2, 66\}$ & \{T1, T6\} & $\{0, 0, 6, 5\}$ & $\{24, 50\}$ \\
$M_{2,25}$ & $\{7793, 7863\}$ & $\{2, 66\}$ & \{T4, T4\} & $\{2, 6, 6, 2\}$ & $\{44, 44\}$ \\
$M_{2,26}$ & $\{7797, 7867\}$ & $\{2, 68\}$ & \{T1, T10\} & $\{0, 0, 6, 12\}$ & $\{24, 60\}$ \\
$M_{2,27}$ & $\{7798, 7868\}$ & $\{2, 68\}$ & \{T2, T6\} & $\{0, 3, 7, 5\}$ & $\{36, 50\}$ \\
$M_{2,28}$ & $\{7799, 7869\}$ & $\{2, 68\}$ & \{T1, T10\} & $\{0, 0, 6, 12\}$ & $\{24, 60\}$ \\
$M_{2,29}$ & $\{7803, 7873\}$ & $\{2, 72\}$ & \{T1, T9\} & $\{0, 0, 6, 8\}$ & $\{24, 56\}$ \\
$M_{2,30}$ & $\{7812, 7882\}$ & $\{2, 76\}$ & \{T1, T7\} & $\{0, 0, 6, 4\}$ & $\{24, 52\}$ \\
$M_{2,31}$ & $\{7813, 7883\}$ & $\{2, 77\}$ & \{T2, T4\} & $\{0, 3, 5, 2\}$ & $\{36, 44\}$ \\
$M_{2,32}$ & $\{7814, 7884\}$ & $\{2, 83\}$ & \{T2, T2\} & $\{0, 3, 3, 0\}$ & $\{36, 36\}$ \\
$M_{2,33}$ & $\{7815, 7885\}$ & $\{2, 86\}$ & \{T1, T6\} & $\{0, 0, 4, 5\}$ & $\{24, 50\}$ \\
$M_{2,34}$ & $\{7816, 7886\}$ & $\{2, 86\}$ & \{T1, T9\} & $\{0, 0, 4, 8\}$ & $\{24, 56\}$ \\
$M_{2,35}$ & $\{7817, 7887\}$ & $\{2, 86\}$ & \{T1, T4\} & $\{0, 0, 4, 2\}$ & $\{24, 44\}$ \\
$M_{2,36}$ & $\{7818, 7888\}$ & $\{2, 86\}$ & \{T1, T9\} & $\{0, 0, 4, 8\}$ & $\{24, 56\}$ \\
\hline
\end{longtable}
\end{center}

\bibliographystyle{JHEP}
\bibliography{reference}

\providecommand{\href}[2]{#2}\begingroup\raggedright\begin{thebibliography}{10}

\bibitem{Candelas:1987kf}
P.~Candelas, A.~M. Dale, C.~A. Lutken and R.~Schimmrigk, \emph{{Complete
  Intersection Calabi-Yau Manifolds}},
  \href{https://doi.org/10.1016/0550-3213(88)90352-5}{\emph{Nucl. Phys. B}
  {\bfseries 298} (1988) 493}.

\bibitem{Anderson:2015iia}
L.~B. Anderson, F.~Apruzzi, X.~Gao, J.~Gray and S.-J. Lee, \emph{{A new
  construction of Calabi\textendash{}Yau manifolds: Generalized CICYs}},
  \href{https://doi.org/10.1016/j.nuclphysb.2016.03.016}{\emph{Nucl. Phys. B}
  {\bfseries 906} (2016) 441--496},
  [\href{https://arxiv.org/abs/1507.03235}{{\ttfamily 1507.03235}}].

\bibitem{Cicoli:2021dhg}
M.~Cicoli, I.~n.~G. Etxebarria, F.~Quevedo, A.~Schachner, P.~Shukla and
  R.~Valandro, \emph{{The Standard Model quiver in de Sitter string
  compactifications}},
  \href{https://doi.org/10.1007/JHEP08(2021)109}{\emph{JHEP} {\bfseries 08}
  (2021) 109}, [\href{https://arxiv.org/abs/2106.11964}{{\ttfamily
  2106.11964}}].

\bibitem{Kreuzer:2000xy}
M.~Kreuzer and H.~Skarke, \emph{{Complete classification of reflexive polyhedra
  in four-dimensions}},
  \href{https://doi.org/10.4310/ATMP.2000.v4.n6.a2}{\emph{Adv. Theor. Math.
  Phys.} {\bfseries 4} (2000) 1209--1230},
  [\href{https://arxiv.org/abs/hep-th/0002240}{{\ttfamily hep-th/0002240}}].

\bibitem{Anderson:2017aux}
L.~B. Anderson, X.~Gao, J.~Gray and S.-J. Lee, \emph{{Fibrations in CICY
  Threefolds}}, \href{https://doi.org/10.1007/JHEP10(2017)077}{\emph{JHEP}
  {\bfseries 10} (2017) 077},
  [\href{https://arxiv.org/abs/1708.07907}{{\ttfamily 1708.07907}}].

\bibitem{Gopakumar:1998ii}
R.~Gopakumar and C.~Vafa, \emph{{M theory and topological strings. 1.}},
  \href{https://arxiv.org/abs/hep-th/9809187}{{\ttfamily hep-th/9809187}}.

\bibitem{Gopakumar:1998jq}
R.~Gopakumar and C.~Vafa, \emph{{M theory and topological strings. 2.}},
  \href{https://arxiv.org/abs/hep-th/9812127}{{\ttfamily hep-th/9812127}}.

\bibitem{Hosono:1994ax}
S.~Hosono, A.~Klemm, S.~Theisen and S.-T. Yau, \emph{{Mirror symmetry, mirror
  map and applications to complete intersection Calabi-Yau spaces}},
  \href{https://doi.org/10.1016/0550-3213(94)00440-P}{\emph{Nucl. Phys. B}
  {\bfseries 433} (1995) 501--554},
  [\href{https://arxiv.org/abs/hep-th/9406055}{{\ttfamily hep-th/9406055}}].

\bibitem{Carta:2021sms}
F.~Carta, A.~Mininno, N.~Righi and A.~Westphal, \emph{{Gopakumar-Vafa
  hierarchies in winding inflation and uplifts}},
  \href{https://doi.org/10.1007/JHEP05(2021)271}{\emph{JHEP} {\bfseries 05}
  (2021) 271}, [\href{https://arxiv.org/abs/2101.07272}{{\ttfamily
  2101.07272}}].

\bibitem{Altman:2014bfa}
R.~Altman, J.~Gray, Y.-H. He, V.~Jejjala and B.~D. Nelson, \emph{{A Calabi-Yau
  Database: Threefolds Constructed from the Kreuzer-Skarke List}},
  \href{https://doi.org/10.1007/JHEP02(2015)158}{\emph{JHEP} {\bfseries 02}
  (2015) 158}, [\href{https://arxiv.org/abs/1411.1418}{{\ttfamily 1411.1418}}].

\bibitem{Altman:2021pyc}
R.~Altman, J.~Carifio, X.~Gao and B.~Nelson, \emph{{Orientifold Calabi-Yau
  Threefolds with Divisor Involutions and String Landscape}},
  \href{https://arxiv.org/abs/2111.03078}{{\ttfamily 2111.03078}}.

\bibitem{Gao:2021xbs}
X.~Gao and H.~Zou, \emph{{Machine Learning to the Orientifold Calabi-Yau with
  String Vacua}},  \href{https://arxiv.org/abs/2112.04950}{{\ttfamily
  2112.04950}}.

\bibitem{Cremmer:1983bf}
E.~Cremmer, S.~Ferrara, C.~Kounnas and D.~V. Nanopoulos, \emph{{Naturally
  Vanishing Cosmological Constant in N=1 Supergravity}},
  \href{https://doi.org/10.1016/0370-2693(83)90106-5}{\emph{Phys. Lett. B}
  {\bfseries 133} (1983) 61}.

\bibitem{Giddings:2001yu}
S.~B. Giddings, S.~Kachru and J.~Polchinski, \emph{{Hierarchies from fluxes in
  string compactifications}},
  \href{https://doi.org/10.1103/PhysRevD.66.106006}{\emph{Phys. Rev. D}
  {\bfseries 66} (2002) 106006},
  [\href{https://arxiv.org/abs/hep-th/0105097}{{\ttfamily hep-th/0105097}}].

\bibitem{Witten:1996bn}
E.~Witten, \emph{{Nonperturbative superpotentials in string theory}},
  \href{https://doi.org/10.1016/0550-3213(96)00283-0}{\emph{Nucl. Phys. B}
  {\bfseries 474} (1996) 343--360},
  [\href{https://arxiv.org/abs/hep-th/9604030}{{\ttfamily hep-th/9604030}}].

\bibitem{Balasubramanian:2005zx}
V.~Balasubramanian, P.~Berglund, J.~P. Conlon and F.~Quevedo,
  \emph{{Systematics of moduli stabilisation in Calabi-Yau flux
  compactifications}},
  \href{https://doi.org/10.1088/1126-6708/2005/03/007}{\emph{JHEP} {\bfseries
  03} (2005) 007}, [\href{https://arxiv.org/abs/hep-th/0502058}{{\ttfamily
  hep-th/0502058}}].

\bibitem{Bianchi:2011qh}
M.~Bianchi, A.~Collinucci and L.~Martucci, \emph{{Magnetized E3-brane
  instantons in F-theory}},
  \href{https://doi.org/10.1007/JHEP12(2011)045}{\emph{JHEP} {\bfseries 12}
  (2011) 045}, [\href{https://arxiv.org/abs/1107.3732}{{\ttfamily 1107.3732}}].

\bibitem{Bianchi:2012pn}
M.~Bianchi, A.~Collinucci and L.~Martucci, \emph{{Freezing E3-brane instantons
  with fluxes}}, \href{https://doi.org/10.1002/prop.201200030}{\emph{Fortsch.
  Phys.} {\bfseries 60} (2012) 914--920},
  [\href{https://arxiv.org/abs/1202.5045}{{\ttfamily 1202.5045}}].

\bibitem{Gao:2013pra}
X.~Gao and P.~Shukla, \emph{{On Classifying the Divisor Involutions in
  Calabi-Yau Threefolds}},
  \href{https://doi.org/10.1007/JHEP11(2013)170}{\emph{JHEP} {\bfseries 1311}
  (2013) 170}, [\href{https://arxiv.org/abs/1307.1139}{{\ttfamily 1307.1139}}].

\bibitem{Cicoli:2012vw}
M.~Cicoli, S.~Krippendorf, C.~Mayrhofer, F.~Quevedo and R.~Valandro,
  \emph{{D-Branes at del Pezzo Singularities: Global Embedding and Moduli
  Stabilisation}}, \href{https://doi.org/10.1007/JHEP09(2012)019}{\emph{JHEP}
  {\bfseries 09} (2012) 019},
  [\href{https://arxiv.org/abs/1206.5237}{{\ttfamily 1206.5237}}].

\bibitem{Cicoli:2013cha}
M.~Cicoli, D.~Klevers, S.~Krippendorf, C.~Mayrhofer, F.~Quevedo and
  R.~Valandro, \emph{{Explicit de Sitter Flux Vacua for Global String Models
  with Chiral Matter}},
  \href{https://doi.org/10.1007/JHEP05(2014)001}{\emph{JHEP} {\bfseries 05}
  (2014) 001}, [\href{https://arxiv.org/abs/1312.0014}{{\ttfamily 1312.0014}}].

\bibitem{Cicoli:2013zha}
M.~Cicoli, S.~Krippendorf, C.~Mayrhofer, F.~Quevedo and R.~Valandro, \emph{{The
  Web of D-branes at Singularities in Compact Calabi-Yau Manifolds}},
  \href{https://doi.org/10.1007/JHEP05(2013)114}{\emph{JHEP} {\bfseries 05}
  (2013) 114}, [\href{https://arxiv.org/abs/1304.2771}{{\ttfamily 1304.2771}}].

\bibitem{Cicoli:2013mpa}
M.~Cicoli, S.~Krippendorf, C.~Mayrhofer, F.~Quevedo and R.~Valandro,
  \emph{{D3/D7 Branes at Singularities: Constraints from Global Embedding and
  Moduli Stabilisation}},
  \href{https://doi.org/10.1007/JHEP07(2013)150}{\emph{JHEP} {\bfseries 07}
  (2013) 150}, [\href{https://arxiv.org/abs/1304.0022}{{\ttfamily 1304.0022}}].

\bibitem{Hebecker:2018yxs}
A.~Hebecker, S.~Leonhardt, J.~Moritz and A.~Westphal, \emph{{Thraxions:
  Ultralight Throat Axions}},
  \href{https://doi.org/10.1007/JHEP04(2019)158}{\emph{JHEP} {\bfseries 04}
  (2019) 158}, [\href{https://arxiv.org/abs/1812.03999}{{\ttfamily
  1812.03999}}].

\bibitem{Cicoli:2021tzt}
M.~Cicoli, A.~Schachner and P.~Shukla, \emph{{Systematics of type IIB moduli
  stabilisation with odd axions}},
  \href{https://arxiv.org/abs/2109.14624}{{\ttfamily 2109.14624}}.

\bibitem{Cicoli:2021gss}
M.~Cicoli, V.~Guidetti, N.~Righi and A.~Westphal, \emph{{Fuzzy Dark Matter
  Candidates from String Theory}},
  \href{https://arxiv.org/abs/2110.02964}{{\ttfamily 2110.02964}}.

\bibitem{Carta:2021uwv}
F.~Carta, A.~Mininno, N.~Righi and A.~Westphal, \emph{{Thraxions: Towards Full
  String Models}},  \href{https://arxiv.org/abs/2110.02963}{{\ttfamily
  2110.02963}}.

\bibitem{Gao:2013rra}
X.~Gao and P.~Shukla, \emph{{F-term Stabilization of Odd Axions in LARGE Volume
  Scenario}},
  \href{https://doi.org/10.1016/j.nuclphysb.2013.11.015}{\emph{Nucl.Phys.}
  {\bfseries B878} (2014) 269--294},
  [\href{https://arxiv.org/abs/1307.1141}{{\ttfamily 1307.1141}}].

\bibitem{Gao:2014uha}
X.~Gao, T.~Li and P.~Shukla, \emph{{Combining Universal and Odd RR Axions for
  Aligned Natural Inflation}},
  \href{https://doi.org/10.1088/1475-7516/2014/10/048}{\emph{JCAP} {\bfseries
  1410} (2014) 048}, [\href{https://arxiv.org/abs/1406.0341}{{\ttfamily
  1406.0341}}].

\bibitem{Bobkov:2010rf}
K.~Bobkov, V.~Braun, P.~Kumar and S.~Raby, \emph{{Stabilizing All Kahler Moduli
  in Type IIB Orientifolds}},
  \href{https://doi.org/10.1007/JHEP12(2010)056}{\emph{JHEP} {\bfseries 12}
  (2010) 056}, [\href{https://arxiv.org/abs/1003.1982}{{\ttfamily 1003.1982}}].

\bibitem{Louis:2012nb}
J.~Louis, M.~Rummel, R.~Valandro and A.~Westphal, \emph{{Building an explicit
  de Sitter}}, \href{https://doi.org/10.1007/JHEP10(2012)163}{\emph{JHEP}
  {\bfseries 10} (2012) 163},
  [\href{https://arxiv.org/abs/1208.3208}{{\ttfamily 1208.3208}}].

\bibitem{Blumenhagen:2010pv}
R.~Blumenhagen, B.~Jurke, T.~Rahn and H.~Roschy, \emph{{Cohomology of Line
  Bundles: A Computational Algorithm}},
  \href{https://doi.org/10.1063/1.3501132, 10.1063/1.3523343}{\emph{J. Math.
  Phys.} {\bfseries 51} (2010) 103525},
  [\href{https://arxiv.org/abs/1003.5217}{{\ttfamily 1003.5217}}].

\bibitem{Blumenhagen:2011xn}
R.~Blumenhagen, B.~Jurke and T.~Rahn, \emph{{Computational Tools for Cohomology
  of Toric Varieties}}, \href{https://doi.org/10.1155/2011/152749}{\emph{Adv.
  High Energy Phys.} {\bfseries 2011} (2011) 152749},
  [\href{https://arxiv.org/abs/1104.1187}{{\ttfamily 1104.1187}}].

\bibitem{He:2018jtw}
Y.-H. He, \emph{{The Calabi\textendash{}Yau Landscape: From Geometry, to
  Physics, to Machine Learning}}.
\newblock Lecture Notes in Mathematics. 5, 2021,
  \href{https://doi.org/10.1007/978-3-030-77562-9}{10.1007/978-3-030-77562-9}.

\bibitem{Constantin:2021for}
A.~Constantin, T.~R. Harvey and A.~Lukas, \emph{{Heterotic String Model
  Building with Monad Bundles and Reinforcement Learning}},
  \href{https://arxiv.org/abs/2108.07316}{{\ttfamily 2108.07316}}.

\bibitem{Abel:2021ddu}
S.~Abel, A.~Constantin, T.~R. Harvey and A.~Lukas, \emph{{String Model
  Building, Reinforcement Learning and Genetic Algorithms}},  in \emph{{Nankai
  Symposium on Mathematical Dialogues}: {In celebration of S.S.Chern's 110th
  anniversary}}, 11, 2021, \href{https://arxiv.org/abs/2111.07333}{{\ttfamily
  2111.07333}}.

\bibitem{Abel:2021rrj}
S.~Abel, A.~Constantin, T.~R. Harvey and A.~Lukas, \emph{{Evolving Heterotic
  Gauge Backgrounds: Genetic Algorithms versus Reinforcement Learning}},
  \href{https://arxiv.org/abs/2110.14029}{{\ttfamily 2110.14029}}.

\bibitem{Larfors:2021pbb}
M.~Larfors, A.~Lukas, F.~Ruehle and R.~Schneider, \emph{{Learning Size and
  Shape of Calabi-Yau Spaces}},
  \href{https://arxiv.org/abs/2111.01436}{{\ttfamily 2111.01436}}.

\bibitem{Blumenhagen:2008zz}
R.~Blumenhagen, V.~Braun, T.~W. Grimm and T.~Weigand, \emph{{GUTs in Type IIB
  Orientifold Compactifications}},
  \href{https://doi.org/10.1016/j.nuclphysb.2009.02.011}{\emph{Nucl.Phys.}
  {\bfseries B815} (2009) 1--94},
  [\href{https://arxiv.org/abs/0811.2936}{{\ttfamily 0811.2936}}].

\bibitem{Collinucci:2008sq}
A.~Collinucci, M.~Kreuzer, C.~Mayrhofer and N.-O. Walliser, \emph{{Four-modulus
  'Swiss Cheese' chiral models}},
  \href{https://doi.org/10.1088/1126-6708/2009/07/074}{\emph{JHEP} {\bfseries
  07} (2009) 074}, [\href{https://arxiv.org/abs/0811.4599}{{\ttfamily
  0811.4599}}].

\bibitem{Cicoli:2016xae}
M.~Cicoli, F.~Muia and P.~Shukla, \emph{{Global Embedding of Fibre Inflation
  Models}}, \href{https://doi.org/10.1007/JHEP11(2016)182}{\emph{JHEP}
  {\bfseries 11} (2016) 182},
  [\href{https://arxiv.org/abs/1611.04612}{{\ttfamily 1611.04612}}].

\bibitem{Cicoli:2018tcq}
M.~Cicoli, D.~Ciupke, C.~Mayrhofer and P.~Shukla, \emph{{A Geometrical Upper
  Bound on the Inflaton Range}},
  \href{https://doi.org/10.1007/JHEP05(2018)001}{\emph{JHEP} {\bfseries 05}
  (2018) 001}, [\href{https://arxiv.org/abs/1801.05434}{{\ttfamily
  1801.05434}}].

\bibitem{Hubsch:1992nu}
T.~Hubsch, \emph{{Calabi-Yau manifolds: A Bestiary for physicists}}.
\newblock World Scientific, Singapore, 1994.

\bibitem{Gorlich:2004qm}
L.~Gorlich, S.~Kachru, P.~K. Tripathy and S.~P. Trivedi, \emph{{Gaugino
  condensation and nonperturbative superpotentials in flux compactifications}},
  \href{https://doi.org/10.1088/1126-6708/2004/12/074}{\emph{JHEP} {\bfseries
  12} (2004) 074}, [\href{https://arxiv.org/abs/hep-th/0407130}{{\ttfamily
  hep-th/0407130}}].

\bibitem{Kallosh:2005yu}
R.~Kallosh and D.~Sorokin, \emph{{Dirac action on M5 and M2 branes with bulk
  fluxes}}, \href{https://doi.org/10.1088/1126-6708/2005/05/005}{\emph{JHEP}
  {\bfseries 05} (2005) 005},
  [\href{https://arxiv.org/abs/hep-th/0501081}{{\ttfamily hep-th/0501081}}].

\bibitem{Tripathy:2005hv}
P.~K. Tripathy and S.~P. Trivedi, \emph{{D3 brane action and fermion zero modes
  in presence of background flux}},
  \href{https://doi.org/10.1088/1126-6708/2005/06/066}{\emph{JHEP} {\bfseries
  06} (2005) 066}, [\href{https://arxiv.org/abs/hep-th/0503072}{{\ttfamily
  hep-th/0503072}}].

\bibitem{Saulina:2005ve}
N.~Saulina, \emph{{Topological constraints on stabilized flux vacua}},
  \href{https://doi.org/10.1016/j.nuclphysb.2005.05.011}{\emph{Nucl. Phys. B}
  {\bfseries 720} (2005) 203--210},
  [\href{https://arxiv.org/abs/hep-th/0503125}{{\ttfamily hep-th/0503125}}].

\bibitem{Kallosh:2005gs}
R.~Kallosh, A.-K. Kashani-Poor and A.~Tomasiello, \emph{{Counting fermionic
  zero modes on M5 with fluxes}},
  \href{https://doi.org/10.1088/1126-6708/2005/06/069}{\emph{JHEP} {\bfseries
  06} (2005) 069}, [\href{https://arxiv.org/abs/hep-th/0503138}{{\ttfamily
  hep-th/0503138}}].

\bibitem{Bergshoeff:2005yp}
E.~Bergshoeff, R.~Kallosh, A.-K. Kashani-Poor, D.~Sorokin and A.~Tomasiello,
  \emph{{An Index for the Dirac operator on D3 branes with background fluxes}},
  \href{https://doi.org/10.1088/1126-6708/2005/10/102}{\emph{JHEP} {\bfseries
  10} (2005) 102}, [\href{https://arxiv.org/abs/hep-th/0507069}{{\ttfamily
  hep-th/0507069}}].

\bibitem{Blumenhagen:2010ja}
R.~Blumenhagen, A.~Collinucci and B.~Jurke, \emph{{On Instanton Effects in
  F-theory}}, \href{https://doi.org/10.1007/JHEP08(2010)079}{\emph{JHEP}
  {\bfseries 08} (2010) 079},
  [\href{https://arxiv.org/abs/1002.1894}{{\ttfamily 1002.1894}}].

\bibitem{Lust:2006zg}
D.~Lust, S.~Reffert, E.~Scheidegger, W.~Schulgin and S.~Stieberger,
  \emph{{Moduli Stabilization in Type IIB Orientifolds (II)}},
  \href{https://doi.org/10.1016/j.nuclphysb.2006.12.017}{\emph{Nucl. Phys.}
  {\bfseries B766} (2007) 178--231},
  [\href{https://arxiv.org/abs/hep-th/0609013}{{\ttfamily hep-th/0609013}}].

\bibitem{Lust:2006zh}
D.~Lust, S.~Reffert, E.~Scheidegger and S.~Stieberger, \emph{{Resolved Toroidal
  Orbifolds and their Orientifolds}},
  \href{https://doi.org/10.4310/ATMP.2008.v12.n1.a2}{\emph{Adv. Theor. Math.
  Phys.} {\bfseries 12} (2008) 67--183},
  [\href{https://arxiv.org/abs/hep-th/0609014}{{\ttfamily hep-th/0609014}}].

\bibitem{Carta:2020ohw}
F.~Carta, J.~Moritz and A.~Westphal, \emph{{A landscape of orientifold vacua}},
  \href{https://doi.org/10.1007/JHEP05(2020)107}{\emph{JHEP} {\bfseries 05}
  (2020) 107}, [\href{https://arxiv.org/abs/2003.04902}{{\ttfamily
  2003.04902}}].

\bibitem{Carta:2021kpk}
F.~Carta, A.~Mininno and P.~Shukla, \emph{{Systematics of perturbatively flat
  flux vacua}}, \href{https://doi.org/10.1007/JHEP02(2022)205}{\emph{JHEP}
  {\bfseries 02} (2022) 205},
  [\href{https://arxiv.org/abs/2112.13863}{{\ttfamily 2112.13863}}].

\bibitem{OGUISO:1993}
K.~Oguiso, \emph{{On Algebraic Fiber Space Structures on a Calabi-Yau 3-Fold}},
  \href{https://doi.org/10.1142/S0129167X93000248}{\emph{International Journal
  of Mathematics} {\bfseries 04} (1993) 439--465}.

\bibitem{Schulz:2004tt}
M.~B. Schulz, \emph{{Calabi-Yau duals of torus orientifolds}},
  \href{https://doi.org/10.1088/1126-6708/2006/05/023}{\emph{JHEP} {\bfseries
  05} (2006) 023}, [\href{https://arxiv.org/abs/hep-th/0412270}{{\ttfamily
  hep-th/0412270}}].

\bibitem{Shukla:2019wfo}
P.~Shukla, \emph{{Dictionary for the type II nongeometric flux
  compactifications}},
  \href{https://doi.org/10.1103/PhysRevD.103.086009}{\emph{Phys. Rev. D}
  {\bfseries 103} (2021) 086009},
  [\href{https://arxiv.org/abs/1909.07391}{{\ttfamily 1909.07391}}].

\bibitem{AbdusSalam:2020ywo}
S.~AbdusSalam, S.~Abel, M.~Cicoli, F.~Quevedo and P.~Shukla, \emph{{A
  systematic approach to K\"ahler moduli stabilisation}},
  \href{https://doi.org/10.1007/JHEP08(2020)047}{\emph{JHEP} {\bfseries 08}
  (2020) 047}, [\href{https://arxiv.org/abs/2005.11329}{{\ttfamily
  2005.11329}}].

\bibitem{Westphal:2006tn}
A.~Westphal, \emph{{de Sitter string vacua from Kahler uplifting}},
  \href{https://doi.org/10.1088/1126-6708/2007/03/102}{\emph{JHEP} {\bfseries
  03} (2007) 102}, [\href{https://arxiv.org/abs/hep-th/0611332}{{\ttfamily
  hep-th/0611332}}].

\bibitem{Kachru:2003aw}
S.~Kachru, R.~Kallosh, A.~D. Linde and S.~P. Trivedi, \emph{{De Sitter vacua in
  string theory}},
  \href{https://doi.org/10.1103/PhysRevD.68.046005}{\emph{Phys. Rev.}
  {\bfseries D68} (2003) 046005},
  [\href{https://arxiv.org/abs/hep-th/0301240}{{\ttfamily hep-th/0301240}}].

\bibitem{Burgess:2003ic}
C.~P. Burgess, R.~Kallosh and F.~Quevedo, \emph{{De Sitter string vacua from
  supersymmetric D terms}},
  \href{https://doi.org/10.1088/1126-6708/2003/10/056}{\emph{JHEP} {\bfseries
  10} (2003) 056}, [\href{https://arxiv.org/abs/hep-th/0309187}{{\ttfamily
  hep-th/0309187}}].

\bibitem{Becker:2002nn}
K.~Becker, M.~Becker, M.~Haack and J.~Louis, \emph{{Supersymmetry breaking and
  alpha-prime corrections to flux induced potentials}},
  \href{https://doi.org/10.1088/1126-6708/2002/06/060}{\emph{JHEP} {\bfseries
  06} (2002) 060}, [\href{https://arxiv.org/abs/hep-th/0204254}{{\ttfamily
  hep-th/0204254}}].

\bibitem{Balasubramanian:2004uy}
V.~Balasubramanian and P.~Berglund, \emph{{Stringy corrections to Kahler
  potentials, SUSY breaking, and the cosmological constant problem}},
  \href{https://doi.org/10.1088/1126-6708/2004/11/085}{\emph{JHEP} {\bfseries
  11} (2004) 085}, [\href{https://arxiv.org/abs/hep-th/0408054}{{\ttfamily
  hep-th/0408054}}].

\bibitem{Rummel:2011cd}
M.~Rummel and A.~Westphal, \emph{{A sufficient condition for de Sitter vacua in
  type IIB string theory}},
  \href{https://doi.org/10.1007/JHEP01(2012)020}{\emph{JHEP} {\bfseries 01}
  (2012) 020}, [\href{https://arxiv.org/abs/1107.2115}{{\ttfamily 1107.2115}}].

\bibitem{Cicoli:2013swa}
M.~Cicoli, J.~P. Conlon, A.~Maharana and F.~Quevedo, \emph{{A Note on the
  Magnitude of the Flux Superpotential}},
  \href{https://doi.org/10.1007/JHEP01(2014)027}{\emph{JHEP} {\bfseries 01}
  (2014) 027}, [\href{https://arxiv.org/abs/1310.6694}{{\ttfamily 1310.6694}}].

\bibitem{Conlon:2012tz}
J.~P. Conlon, \emph{{Quantum Gravity Constraints on Inflation}},
  \href{https://doi.org/10.1088/1475-7516/2012/09/019}{\emph{JCAP} {\bfseries
  09} (2012) 019}, [\href{https://arxiv.org/abs/1203.5476}{{\ttfamily
  1203.5476}}].

\bibitem{Demirtas:2019sip}
M.~Demirtas, M.~Kim, L.~Mcallister and J.~Moritz, \emph{{Vacua with Small Flux
  Superpotential}},
  \href{https://doi.org/10.1103/PhysRevLett.124.211603}{\emph{Phys. Rev. Lett.}
  {\bfseries 124} (2020) 211603},
  [\href{https://arxiv.org/abs/1912.10047}{{\ttfamily 1912.10047}}].

\bibitem{Carta:2201aaaaa}
F.~Carta, A.~Mininno and P.~Shukla, \emph{{Systematics of perturbatively flat
  flux vacua for pCICYs}},  \href{https://arxiv.org/abs/2201.xxxxx}{{\ttfamily
  2201.xxxxx}}.

\end{thebibliography}\endgroup

\end{document}